\documentclass[12pt,a4paper]{article} 
\pdfoutput=1
\usepackage{jheppub}
\usepackage{enumerate}
\usepackage{epsfig} 
\usepackage{float} 
\usepackage[caption = false]{subfig}
\usepackage{wasysym} 
\usepackage{mathrsfs} 
\usepackage{amsfonts} 
\usepackage{amsbsy} 
\usepackage{amscd} 
\usepackage{pstricks} 
\usepackage{multirow} 
\usepackage{tikz}
\usepackage{slashed}
\usetikzlibrary{arrows,positioning,shapes.geometric} 
\usepackage{color}
\definecolor{myred}{rgb}{0.6,0,0} 
\definecolor{myblue}{rgb}{0,0.2,0.4}
\definecolor{mygreen}{rgb}{0,0.9,0.1}
\definecolor{hc}{rgb}{.9,0.1,0.7}
\definecolor{hcout}{rgb}{.9,0.7,0.9}
\definecolor{Orange}{rgb}{1.,0.65,0.}

\numberwithin{equation}{section}
\numberwithin{figure}{section}
\numberwithin{table}{section}

\newcommand{\met}{\ensuremath{\not\!\!E_T}\xspace}
\newcommand{\be}{\begin{equation}}
\newcommand{\ee}{\end{equation}}
\newcommand{\bea}{\begin{eqnarray}}
\newcommand{\eea}{\end{eqnarray}}

\newcommand{\newc}{\newcommand}
\newc{\bi}{\begin{itemize}}
\newc{\ei}{\end{itemize}}

\newc{\ra}{\rightarrow}
\newc{\sq}   {\mbox{$\wt{q}$}}
\newc{\msq}  {\mbox{$m_{\sq}$}}
\newc{\gl}   {\mbox{$\wt{g}$}}
\newc{\mgl}  {\mbox{$m_{\gl}$}}
\def \met  {\mbox{${E\!\!\!\!/_T}$}}
\newc{\wt}{\widetilde}

\def \lspone{\wt\chi_1^0}
\def \mlspone{m_{\lspone}}

\newc{\ifb}{\mbox{${\rm fb}^{-1}$}}

\def \chonepm{\wt\chi_1^\pm}

\newc{\del}{\delta}

\def \lstop{\wt {t}_1}
\def \mlstop{m_{\lstop}}
\def \lsbot{\wt {b}_1}
\def \mlsbot{m_{\lsbot}}

\def \stopL{\wt {t}_L}
\def \stopR{\wt {t}_R}

\title{A revisit to a compressed supersymmetric spectrum with 125 GeV Higgs } 
\author[a]{Juhi Dutta,}  
\author[b]{Partha Konar,}
\author[a]{Subhadeep Mondal,}
\author[a]{Biswarup Mukhopadhyaya}
\author[a]{and Santosh Kumar Rai}
\affiliation[a]{ Regional Centre for Accelerator-based Particle Physics, \\
Harish-Chandra Research Institute, \\ Chhatnag Road, Jhusi, Allahabad 211019, India}
\affiliation[b]{Physical Research Laboratory,  Ahmedabad 380009, India}
\emailAdd{juhidutta@hri.res.in}
\emailAdd{konar@prl.res.in}
\emailAdd{subhadeepmondal@hri.res.in}
\emailAdd{biswarup@hri.res.in}
\emailAdd{skrai@hri.res.in}

\abstract{
A compressed spectrum was initially proposed as an explanation for 
the elusiveness of low-energy supersymmetry (SUSY). Some characteristic signals at 
the Large Hadron Collider (LHC), such as mono-jet +$\met$, had been 
propounded as its trademark signals. However, later investigations suggested that lower 
limits on the supersymmetric particle masses would be quite stringent in spite of 
compression. Also, most compressed SUSY scenarios studied so far are only 
partially compressed. In this backdrop, we make an exhaustive analysis of the 
compressed SUSY scenarios for the 13 TeV run of LHC, keeping the level of compression
in the entire spectrum as high as possible. A broad class of benchmark spectra are thus 
considered, after ensuring consistency with the observed Higgs mass as well as 
the dark matter constraints. The rates of observable events in the high-energy 
run are obtained through detailed simulation, for both the multi-jet +$\met$ and mono-jet + $\met$  
final states. Our conclusion is that the former is still more efficient to reveal a 
compressed SUSY spectrum first, while the latter can serve as a useful 
confirmatory channel.
}

\preprint{HRI-RECAPP-2015-018}

\keywords{Supersymmetry Phenomenology, Large Hadron Collider, Multi-jets}
\begin{document}

\maketitle



\section{Introduction}
\label{sec:intro}

Despite the very pertinent candidature of TeV-scale supersymmetry
(SUSY) as the solution to the Higgs naturalness problem, together with the possibility 
of solving the dark matter (DM) puzzle with its help, the Large Hadron
Collider (LHC) experiment is yet to reveal any hint of SUSY. A way of retaining
one's hope in this direction is to think of some version(s) of SUSY, broken 
around the TeV-scale, but with such spectra as can suppress the usually
expected signals. One such version assumes sparticle masses to be 
compressed within a rather small range, a situation whose
theoretical justification and phenomenological analyses have already generated some 
efforts \cite{LeCompte:2011cn,LeCompte:2011fh,Murayama:2012jh,Nakayama:2013uta}. 
The compressed spectrum causes the jets and leptons produced
in SUSY cascades to be relatively soft, and also downgrades the missing transverse 
energy ($\met$) somewhat, thus potentially suppressing signals that pass the acceptance
criteria at the LHC. One can therefore envision allowed regions in the parameter space
after the 8 TeV run, with relatively low-lying superparticles but
small spacing between the squark/gluino masses and that
of the lightest SUSY particle (LSP)\footnote{The lightest neutralino
($\lspone$) has been assumed to be the LSP in this study.}.

It was initially thought that the best way to look for compressed SUSY
was to focus on the mono-jet +$\met$ signal \cite{Alves:2010za,Alvarez:2012wf,Dreiner:2012gx,Bhattacherjee:2012mz,Dreiner:2012sh,Bhattacherjee:2013wna,
Martin:2008aw,Belanger:2012mk,Cohen:2013xda,TheATLAScollaboration:2013fha,Chatrchyan:2011nd,Aad:2011xw,Aad:2012fqa,
Mukhopadhyay:2014dsa,Khachatryan:2014rra,Aad:2015zva}. Subsequent investigations
in the  context of run-I  showed that
the `conventional' multi-jet $+\met$ signals (with or without accompanying
leptons) could  be more useful if appropriate event selection
criteria were followed \cite{Dreiner:2012gx,Bhattacherjee:2012mz,Bhattacherjee:2013wna}. 
It is  important to see how such multi-jet +$\met$ signals fare against
the mono-jet +$\met$ ones in the 13 / 14 TeV runs of the LHC. 

A few things, however, remain to be noted carefully in such an investigation.
In many recent studies, experimental as well as theoretical, the
deciding factor is assumed to be the mass splitting between the LSP and the coloured members such as
gluino/squarks, the role of the rest of the spectrum being relatively 
inconsequential. It is also sometimes customary to focus on
the mass gap between the LSP and the next-to-LSP (NLSP). This kind of
an approach has often been prompted by attempts to parametrise the
spectrum in terms of some `compression factor'
\cite{LeCompte:2011cn,LeCompte:2011fh} that straightjackets the entire
spectrum in a little oversimplified manner.  However, one should take
an equally serious note of the rest of the minimal SUSY standard model
(MSSM) spectrum where even non-coloured particles (or third family
squarks) can have substantial splitting with the LSP, thus
producing additional hard jets and/or leptons after all.

Another vital issue that needs to be addressed is the undeniable presence of the lighter
CP-even Higgs boson around 125-126 GeV. 
In a SUSY extension of the standard model (SM), one can only consider spectra where this
mass value is replicated, its behaviour being most likely SM-like. As
we know, the mass of this scalar in the MSSM, taking radiative
corrections into account, is highly dependent on the two stop masses
as well as the stop left-right mixing angle. Hence the degree of
compression of the entire MSSM spectra is expected to be strongly
constrained, if the lighter CP-even Higgs mass has to be in the right
value. Therefore, the compressed spectra proposed in the earlier works
need to be revisited in the aftermath of the Higgs boson
discovery. This is not thoroughly done in most existing studies; it is
often implied that either the spectrum is only partly compressed
\cite{Alves:2010za,Alvarez:2012wf,Dreiner:2012gx,Bhattacherjee:2012mz,Dreiner:2012sh,Bhattacherjee:2013wna,
Martin:2008aw,Belanger:2012mk,Cohen:2013xda,TheATLAScollaboration:2013fha,Chatrchyan:2011nd,Aad:2011xw,Aad:2012fqa,
Mukhopadhyay:2014dsa,Khachatryan:2014rra,Aad:2015zva,Chalons:2015vja},
or some physics beyond MSSM is responsible for the observed value of
the Higgs mass \cite{Aad:2012tfa,Chatrchyan:2012xdj,Aad:2015zhl}. In contrast, we have 
proceeded assuming the intervention of only
the MSSM fields in deciding the Higgs mass(es).

In addition, the constraints from the relic density of the universe as
well as those arising from direct DM search experiments are important
requirements of a SUSY spectrum.  We have taken these constraints into
account while selecting the benchmark points in the parameter space.
For more detailed study of DM in the context of compressed SUSY scenario, see 
\cite{Martin:2007gf,Martin:2007hn,Baer:2007uz}.

On the whole, given the manifold diversity of an MSSM spectrum,
{\em we have preferred to think not in terms of some compression
parameter(s) in a somewhat simplified spectrum but to work with
a wide assortment of benchmark points, which  reflect as many different
possibilities as possible.} We have kept the heavier stop mass and/or
the Higgsino mass parameter $\mu$ somewhat above the compressed 
spectrum in some cases. The latter choice may perhaps be justified
by the observation that $\mu$ does not have the same origin as
the SUSY-breaking mass parameters; it is in fact a SUSY-invariant
parameter, though destined to be in the TeV scale by the electroweak
symmetry breaking requirement. In any case we have also presented
results for some benchmark points where the {\em  entire spectrum}  
lies tightly compressed. After a detailed study of this variety of benchmarks,
we reach the conclusion that signals comprising multi-jets are
likely to be more useful in the 13/14 TeV runs, as compared to those
depending upon mono-jets.

In section 2, we discuss the existing experimental limits on the MSSM parameter space. We further discuss the status of compressed SUSY search at the LHC. Then we look for a truly compressed SUSY spectrum keeping the lightest CP-even Higgs boson mass in its allowed range around 125 GeV. While doing so, we carry out a detailed scan of the relevant parameter space keeping all the collider, DM and flavour physics constraints in consideration. We then provide some benchmark points to showcase our results with different squark-gluino mass hierarchy keeping the lightest neutralino as the LSP. In section 3, we explore the collider aspects of such scenarios in the context of run II of the LHC. We look for both multi-jet +$\met$ and mono-jet +$\met$ final states arising from all possible squark-gluino production channels and compare the sensitivities of these two signals to such compressed spectrum and conclude.  

\section{Status of SUSY search and a compressed spectrum}
\label{sec:status}
The generic SUSY search channels at the LHC involve the strongly
interacting sector comprising of squarks ($\sq$) and gluino ($\gl$), all
of which have large production rates. In the CMSSM/mSUGRA scenario,
the mass spectrum for the squarks, gluino and other sparticles have a
predetermined hierarchy dictated by the renormalisation group (RG)
evolutions, once the free parameters are chosen at the unification
scale. Once a mass-ordering is thus established, this simplifies the
search strategies, since the observed jets or charged leptons
originating from the SUSY cascades would carry the imprint of the mass
spectrum.  One usually associates the signal to contain jets and
charged leptons with large transverse momenta along with substantial missing transverse
energy ($\met$) carried away by the stable lightest SUSY particle
(LSP).  As a result, the final states are easily separated from their
respective SM backgrounds and the exclusion limits derived
on the coloured sparticles come out stronger in this framework.  Both
CMS and ATLAS have put bounds which are close to around 1 TeV on the
squark masses and 1.4 TeV on the gluino masses respectively for simplified models
\cite{Aad:2014lra}. In the case of degenerate squarks and gluinos, the
exclusion limit extends upto 1.7 TeV in CMSSM models \cite{Aad:2014wea}. 

\subsection{Current limits on MSSM from ATLAS and CMS}
\label{sec:status:mssm}
However, the MSSM in general poses a bigger
challenge for LHC to put similar exclusion limits. Since the number of
free parameters increases manifold, possibilities for different mass
ordering of the SUSY particles open up. In such situations, it
not only becomes very difficult to put absolute bounds on the masses of
the sparticles, but the guiding principles to search for SUSY at LHC also
become ambiguous.  Because of this, the bounds are always associated with
some simplified assumptions for the decay pattern of the produced
particles and therefore, one has to be careful while implementing
these limits. In  such scenarios, gluino mass ($\mgl$) is
excluded upto 1.3 - 1.5 TeV when the lightest neutralino (LSP) mass
($\mlspone$) is not heavier than 500 - 600 GeV \cite{Aad:2014wea}, provided the
first two generation squarks are lighter than gluino. When the squarks
are much heavier than the  gluino, the $\gl$ decays via off-shell
squarks. The decay to three-body final state comprising of two quarks
and the LSP leads to softer jets in the final state which dilute the
$\mgl$ exclusion limit to about 1.4 TeV for $\mlspone\le 300$ GeV
\cite{Aad:2014wea}.  Just as above, all such available limits from
run-I data of the LHC are expected to weaken further if the mass
difference between the parent and daughter particles gets reduced as
this would result in less $\met$ and softer jets/leptons in the final
state. For example, if $\mgl - \mlspone$ is reasonably small, the exclusion
limit on $\mgl$ reduces to 550-600 GeV \cite{Aad:2014wea}. 
Thus, a light spectrum with small mass gaps among the SUSY particles might 
have escaped run-I scrutiny, thereby prompting increased interest in 
a {\it{compressed SUSY scenario}} \cite{LeCompte:2011cn,LeCompte:2011fh}.

Summarising the other available bounds on MSSM, for a much heavier
gluino, lighter squark (first 2 generations) masses are excluded below
850 GeV when $\mlspone\le 350$ GeV \cite{Aad:2014wea}.  Lighter stop
masses ($\mlstop$) are excluded upto 600-700 GeV provided $\lstop$
decays into a top quark ($t$) and $\lspone$ where $\mlspone < 250$ GeV
\cite{Aad:2014kra,Aad:2014bva}. When the $\lstop$ decays into a bottom
quark ($b$) and the lighter chargino ($\chonepm$), any $\mlstop$ below
500-600 GeV is excluded for $\mlspone$ below 200-250 GeV
\cite{Aad:2014kra,Aad:2014qaa}, the exact limits being dependent on
the chargino mass. For other decay modes of $\lstop$ (flavour
violating or $> 2$-body modes), the exclusion limits reduce to 240-260
GeV \cite{Aad:2014kra,Aad:2014qaa,Aad:2014nra}. Similarly, a lighter
sbottom mass ($\mlsbot$) below 620 GeV  is excluded for
$\mlspone\le 150$ GeV \cite{Aad:2013ija}. When $\mlsbot - \mlspone$ is
small, the exclusion limit on $\mlsbot$ is lowered to 250 GeV
\cite{Aad:2014nra}.

Since for our present work we consider a relatively compressed
spectrum, it turns out that the weakly interacting sector of MSSM has
a relatively less important role to play. Therefore, we shall focus 
on the production and decay of the coloured sparticles. For
a recent summary of SUSY search limits at the LHC, we refer the readers, for example, to
Ref.~\cite{CMSSUSY,ATLASSUSY}.

\subsection{SUSY with the entire spectrum compressed}
\label{sec:status:compressed}
Compressed SUSY spectra has been studied in the context of LHC quite
extensively with special emphasis on the smallness of the mass
gap between the coloured sparticles and the LSP. A coloured NLSP (be
it a squark or a gluino) is often assumed, and the role of the other
sparticles in SUSY signals is considered to be of secondary
importance.  
In an un-compressed spectrum one probably can accept that the significant 
contribution to the rates come from lightest coloured sparticle production 
\cite{Konar:2010bi}
(where the other coloured modes are heavier).  
Understandably, hard jets or leptons are difficult to
obtain in the final state for small mass gaps.  This results in weaker
limits on the parameter space, when compared to the standard SUSY
searches. However, such effects do not always presume the entire
spectrum to be compressed.  In most cases, a part of the strongly
interacting sector (for example, the third family squarks) is ignored
by decoupling it from the low lying spectrum.  In addition, many
extant studies do not pay enough attention to parts of the coloured
spectrum, which may not be entirely decoupled, but whose participation
vis-a-vis that of the gluino may have bearing on the SUSY signals,
especially on the kinematic profiling of the events arising out of
sparticle production. For example, the contribution to the final state
may dominantly come from the hard processes comprising of the
production of squarks in association with gluinos. Now, inspite of
having a small gluino-LSP mass gap, the squarks may have a substantial
mass gap with the LSP.  These sparticles will then start contributing
to the final state giving rise to harder jets or leptons along with
relatively larger $\met$. Hence the question we really need to ask is, 
how would a really compressed
SUSY spectrum, with almost all sparticles rubbing shoulders with each
other, play out at the LHC. 

Such a SUSY spectrum, however, has to obey some guiding
principles. The first of these is to reproduce the lighter
neutral CP-even Higgs mass in the neighbourhood of 125 GeV.
The next constraint to be taken into account is the contribution
to the relic density of the universe. These, in addition to 
various limits arising from flavour physics and/or direct search results
till date, guide one towards some allowed spectra that are either
fully compressed or have to leave out some relatively heavy states above
the compressed band. 

We discuss these issues next, based on which we
finally choose specific benchmarks from the viable parameter space
that highlight different mass hierarchies among the gluino and the
squark states. We use the benchmarks to carry out a detailed collider
simulation for both multi-jet +$\met$ and mono-jet +$\met$ final states, to determine
which search strategy may help us better to discover or rule out
various SUSY spectra that are compressed to the utmost.
\subsection{A spectrum constrained by Higgs mass and dark matter}
\label{sec:status:constraint}
We recall that the tree-level mass of the lightest CP-even Higgs boson as obtained in the MSSM framework 
has an upper bound:
\bea
m_h^{\tt tree} \leq m_Z ~| \cos 2\beta|
\label{Higgs_tree}
\eea 
where $\tan\beta = v_u/v_d$ is the ratio of the two Higgs VEVs.  Since
Eq.~\ref{Higgs_tree} cannot allow a Higgs mass greater than the $Z$
boson mass, one has to rely on substantial contribution through higher
order (loop) corrections to reach the neighbourhood of 125 GeV,
the experimentally measured mass of what could be the lighter CP-even 
neutral scalar in a SUSY scenario. The dominant higher-order contribution 
comes from stops  in the loop due to a large Yukawa
coupling of the Higgs boson with the top quark.  The one-loop
contribution to the $m_h^{\tt tree}$ is approximately \cite{Hall:2011aa}:
\bea (\Delta m_h^2)^{\tt 1-loop}\simeq\frac{3m_t^4}{4\pi^2v^2}\left ({\rm
  ln}\frac{M_{S}^2}{m_t^2}+\frac{X_t^2}{M_{S}^2}-
\frac{X_t^4}{12M_{S}^2}\right ),
\label{Higgs_corrcn}
\eea where $v$ is the up-type Higgs VEV,
$M_{S}=\sqrt{m_{\stopL}m_{\stopR}}$ is the geometric mean of the stop
left-right masses and $X_t=A_t-\mu \cot\beta$, which governs
$\stopL-\stopR$ mixing as well as the splitting between the two stop
mass eigenstates.  Thus the radiatively corrected Higgs mass crucially
depends on two parameters, namely, $M_{S}$ and $X_t$, along with $\mu$
and $\tan\beta$.  We note that in order to have one of the CP-even
Higgs mass as 125 GeV Higgs boson in the theory, one requires large
stop masses and large stop mixing ($X_t\simeq\pm\sqrt{6}M_{S}$)
\cite{Djouadi:2015bba,Carena:2013ytb}. 

One has the freedom to choose soft-breaking SUSY parameters in the
MSSM for each sfermion generation separately. Also, maximum mass
splitting is possible in the third family (due to the
larger Yukawa couplings) which again contributes most significantly
to the Higgs mass correction. Thus one concludes that obtaining a
significant compression in the {\it entire spectrum} is difficult, 
since achieving  $m_h \approx$ 125 GeV requires (at least) one stop 
eigenstate to be heavy. 

At the same time, we find a somewhat large $\mu$, too, is 
favourable in achieving $m_h \approx$ 125 GeV. However, this also
entails the possibility of having the Higgsino-dominated chargino and 
neutralinos on the heavier side, thus jeopardising the degree
of compression in the entire MSSM spectrum. This also
affects the Higgsino component in the LSP, which in turn 
may reduce the annihilation rate far too much, leading to
excess relic density. 

We thus use the following constraints in our scan of the parameter space :
\begin{itemize}
\item The lightest CP-even  Higgs mass should  be in the range $122 < m_h < 128$ GeV 
\cite{Aad:2012tfa,Chatrchyan:2012xdj,Aad:2015zhl}.
\item The LEP lower bound on the lightest chargino mass,
  \emph{viz.} $m_{\widetilde\chi^{\pm}_1} > 103.5$ GeV
  \cite{LEPSUSYWG}.
\item Constraints from branching ratios of rare decays such as
  $\rm{BR}(b\rightarrow s\gamma)$ and
  $\rm{BR}(B_s\rightarrow\mu\mu)$ \cite{Aaij:2013aka,Chatrchyan:2013bka}.
\item The LSP, $\lspone$, which is the cold dark matter candidate, satisfies the observed thermal relic density, 
0.092 $< \Omega_{\wt\chi} h^2 <$ 0.138 \cite{Hinshaw:2012aka}.

However, for our parameter scan we have considered only the upper
limit of $\Omega h^2$, taking the view that it is plausible to have 
multi-component DM~\cite{Berezhiani:1989fp,Boehm:2003ha,Ma:2006uv,
Gu:2007gy,Hur:2007ur,Cao:2007fy,Hambye:2008bq,Aoki:2012ub,Heeck:2012bz,Konar:2009qr}.
However, substantial portions in the parameter space has been identified, 
where a single-component DM satisfies. 
We also include the constraints from direct dark matter searches, as obtained from the LUX data 
\cite{Akerib:2013tjd}.  
\end{itemize}

In order to achieve spectra which are as compressed as can be,
consistent with the above constraints, we have taken into consideration the following points 
in our prediction of the LHC signal: 
\begin{itemize}
\item The mass gap within the stop pair being large, overall compression can be reduced in
situations where one stop eigenstate, $\lstop$, lies just above the neutralino LSP.
\item Gluino can be light and both cases are considered when gluino mass is above or below the lighter stop. 
\item The non-strongly interacting sfermions and gauginos
are assigned various orders in the compressed spectrum. Though they
have less of a role in the LHC signals, they may 
have a bearing on the relic density as well as cascade decays. 
\item The heavier stop mass as well as $\mu$ are kept both outside
and inside the most compressed part of the spectrum. The latter possibility ({\it i.e.} no
sparticle outside the compressed region) works for relatively
heavy spectra only.
\end{itemize}

\begin{figure}[t]
\centering
\includegraphics[scale=0.45]{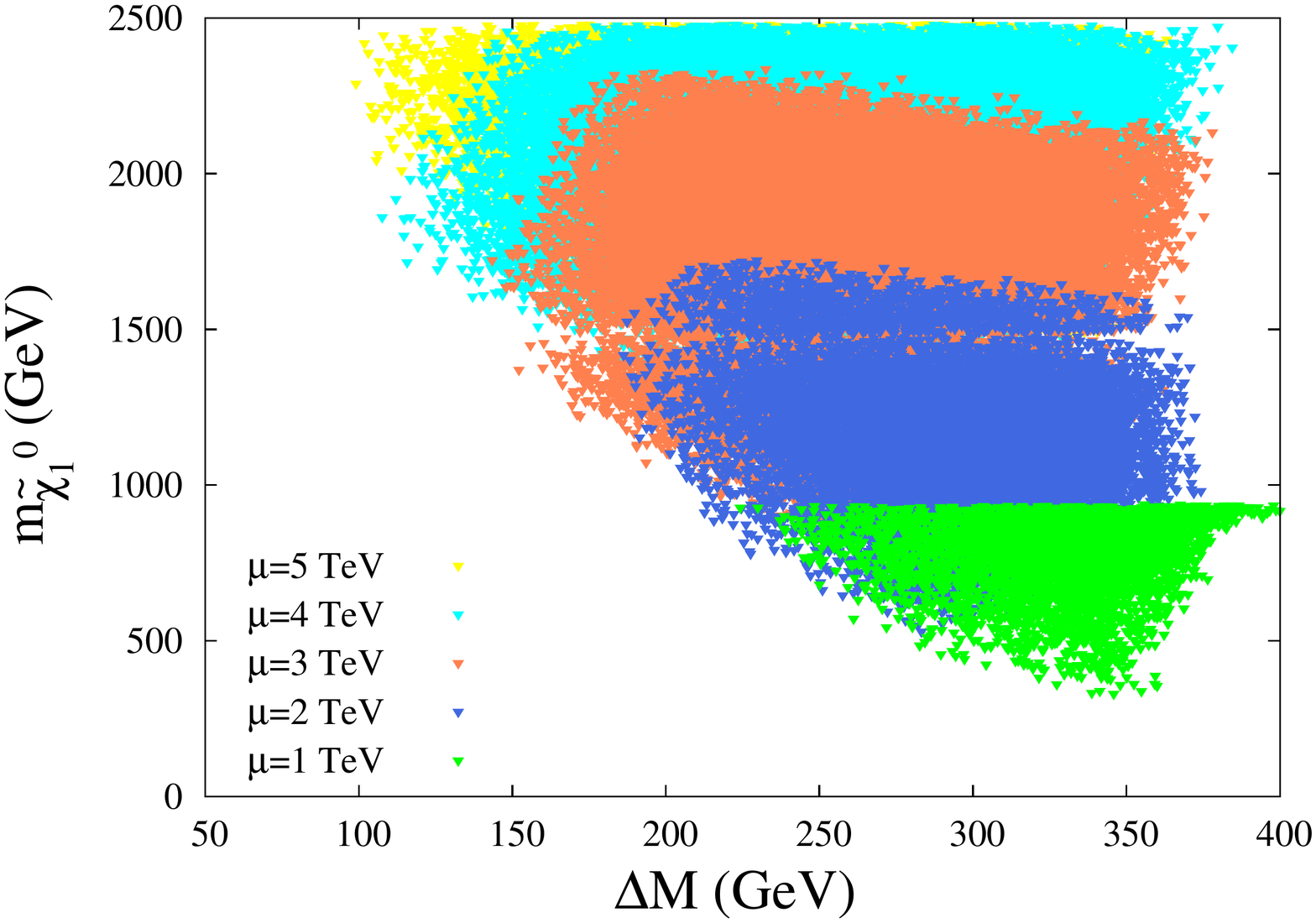}
\caption{Distribution of $\mlspone$ as a function of $\Delta M$ at different $\mu$ values. 
The five colours (yellow, cyan, brown, blue and green) indicate five different values of $\mu$. 
The points in the plot satisfy all the relevant constraints mentioned in the text.}
\label{fig:dm_mlsp}
\end{figure}

We parameterise the compression using the mass gap
between the LSP ($\mlspone$) and the heaviest 
 sparticle ($\wt X$) in the spectrum,
defined as $\Delta M = m_{\wt X} - \mlspone$, where $\wt X \in$ [$\gl$, $\widetilde t_2$, $\widetilde b_2$,
$\widetilde\tau_2$, $\widetilde\chi_2^0$, $\widetilde\chi_1^{\pm}$]\footnote{Note that the higgsino dominated states may lie outside 
our compressed spectrum when $\mu$ is chosen to be very large.}.  
We scan over the relevant parameters shown in Table~\ref{tab:scan_range}. 
\begin{table}[H]
\centering
\begin{tabular}{||c|c||} \hline\hline
Parameters  & Ranges \\
\hline\hline 
$M_1$, $M_2$, $M_3$ & (100, 2500) GeV \\
$A_t$ & (-3000, 3000) GeV \\
$\tan\beta$ & (2, 50) \\
$M_L$ = $M_R$ & $(M_1, M_1 + 200) ~{\rm GeV} (\rm if ~M_1 < M_2)$ \\ 
              & $(M_2, M_2 + 200) ~{\rm GeV} (\rm if ~M_2 < M_1)$ \\   
\hline\hline
\end{tabular}
\caption{Ranges of the relevant parameters for the scan. $M_1$, $M_2$, $M_3$
are the gaugino mass parameters, varied in the same range but independent of 
each other. $M_L$ and $M_R$ are the left-handed and the right-handed soft 
mass parameters of squarks and sleptons.}
\label{tab:scan_range}
\end{table}
Here $M_L$ and $M_R$ represent the soft mass parameters of the
left and right handed squarks and sleptons respectively\footnote{Although the soft mass parameters for the squarks and sleptons are kept equal by choice, this does not significantly affect the hadronic signals.}.
Table~\ref{tab:scan_range} suggests, we chose same $M_L$ and 
$M_R$ for all flavours. For our scan, we have used {\tt SPheno}(v3.3.6) \cite{Porod:2003um,Porod:2011nf} 
which calculates all sparticle masses at one-loop level while the Higgs mass is 
calculated at two-loop in order 
to generate the SUSY spectrum and consequently {\tt micrOMEGAs}(v4.1.7) \cite{Belanger:2013oya} to 
calculate the DM relic density and direct-detection cross-section, 
flavour physics constraints and muon g-2. 
In Fig.~\ref{fig:dm_mlsp}, we plot LSP mass $\mlspone$ as a function of compression mass gap $\Delta
M$. As evident, a $\mu$-value close to or above 4 TeV allows a $\Delta M$
as low as 100 GeV. This figure gives a clear idea of the heaviness of
the MSSM spectra as we keep compressing the whole spectrum. 
To give some estimate, in order to restrict the spectrum with $\Delta M\sim 100$ GeV, 
one obtains a lower limit on the LSP mass close to 1800 GeV for $\mu=5$ TeV.

We examine next how the the constraints from relic density($\Omega h^2$) 
and the spin-independent cross-sections 
($\sigma_{SI}$) in direct search experiments affect the allowed parameter space. Since we are 
considering a compressed MSSM scenario, 
there are always some sparticles whose masses lie close enough to the LSP 
to produce sufficient co-annihilation 
to bring down the relic density to permissible limits. For a wino-like 
LSP, the $\lspone$ mainly co-annihilates 
with the $\chonepm$. In addition, if there are sparticles nearby, e.g, 
$\gl$, $\lstop$, $\lsbot$, $\widetilde\tau_1$, 
in the spectrum, all the annihilation channels combine to produce 
underabundance of the DM relic density. Similar 
situation may occur in case of a bino-like or a bino-wino mixed LSP 
state. Hence $\Omega h^2$ is not a very serious 
constraint for such a scenario. 

\begin{figure}[t]
\centering
\includegraphics[scale=0.45]{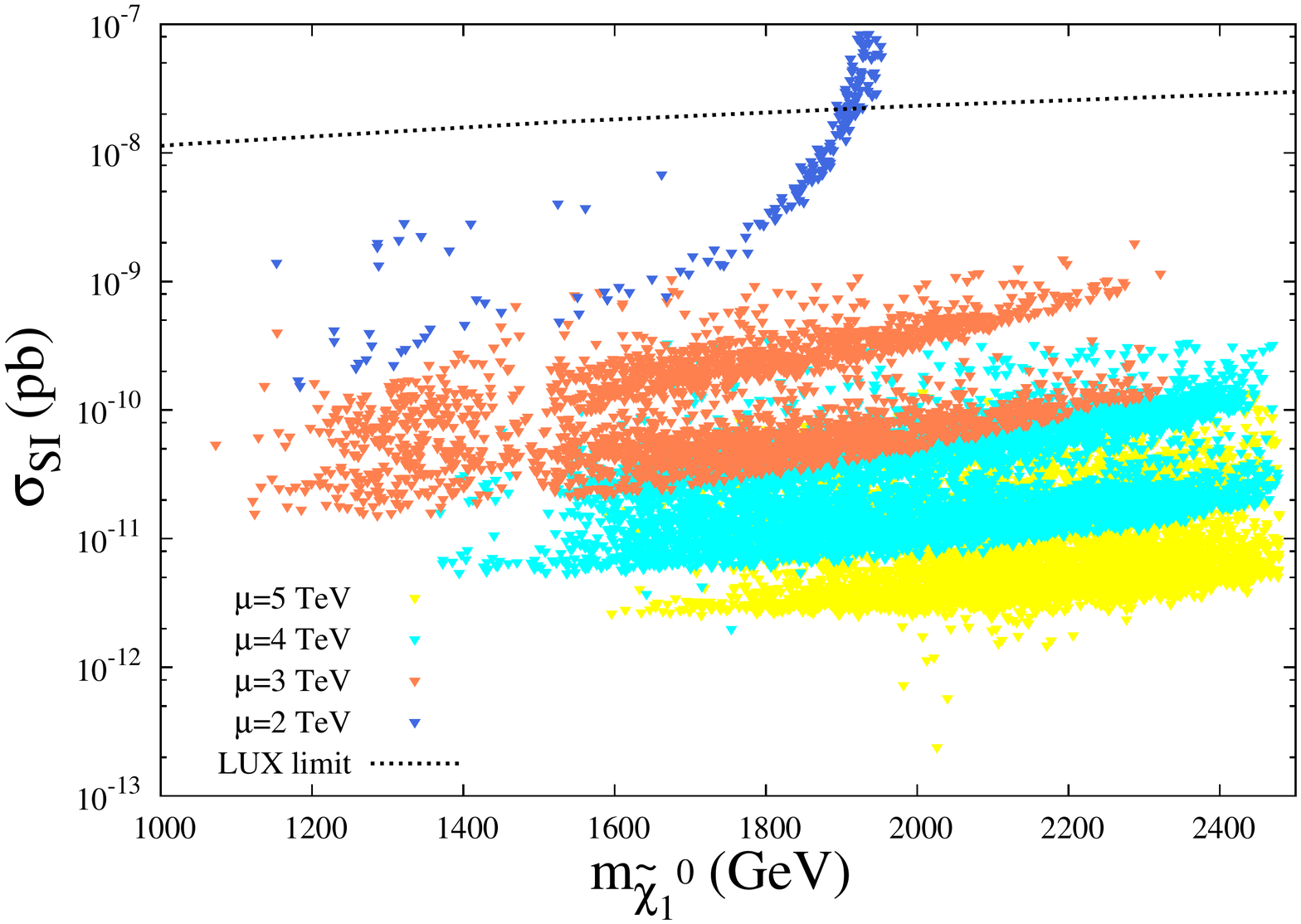}
\caption{The direct detection cross-section as a function of the LSP mass. 
Since we are interested in small $\Delta M$ we have plotted the points only when  
$\Delta M \leq 200$ GeV. Colour labels follow Fig.~\ref{fig:dm_mlsp}. 
The black dotted line represents the most updated LUX bound.}
\label{fig:sigsi_mlsp}
\end{figure}
Direct search limits, ($\sigma_{SI}$), however, can rule out 
some of the relevant parameter space. 
Fig.~\ref{fig:sigsi_mlsp} shows the distribution of $\sigma_{SI}$ as a 
function of the DM mass($m_{\wt\chi_1^0}$).  
Note that in this plot we only show those points in the parameter space, 
which produce $\Delta M \leq 200$ GeV. Understandably, there are no points 
corresponding to $\mu=1$ TeV in the distribution, 
since Fig.~\ref{fig:dm_mlsp} clearly shows the maximum compression we can 
reach in this case is close to 220 GeV. 
The black dotted line represents the most recent bound on $\sigma_{SI}$ provided by 
the LUX experiment as a function of the DM mass \cite{Akerib:2013tjd}. As expected, 
all the points obtained in the scan with $\mu \geq 3$ TeV lie well below the 
exclusion line, the LSP in these scenarios have almost zero contribution from higgsino components 
and as a result, the Z-boson coupling of the LSP
is reduced to a very small value, resulting in such small DM-nucleon scattering cross-sections. 
However, if we keep decreasing the $\mu$ value, $\sigma_{SI}$ increases. When the 
bino or wino mass parameters become comparable 
to the $\mu$ parameter, as happens in part of the parameter space in the $\mu=2$ TeV 
case\footnote{This is a result of our choice of the scan ranges of $M_1$ and $M_2$ as indicated 
in Table~\ref{tab:scan_range}. In Sec.~\ref{sec:status:benchmark}, we show two such sample benchmark 
points with non-negligible higgsino component (e.g. 8\% in {\tt BP6}). However, we have not considered 
higgsino-like LSP for our present work.}, the
LSP turns out to be a mixed state with 
substantial higgsino component. This results in enhancement of $\sigma_{SI}$ which 
is manifested in the few blue points in the figure which violate the LUX limit.

To demonstrate how the stop mixing parameters behave 
under the Higgs mass constraint, we chose one particular LSP mass  
close to 1100 GeV ($M_1=1100$ GeV) and vary $A_t$ in the range $(-3000, 3000)$ GeV and 
$\wt t_L$ and $\wt t_R$ soft masses, $M^{Q_3}_L$ and $M^{U_3}_R$, such that
$M_1 < M^{Q_3}_L (M^{U_3}_R) < M_1 + 200$ GeV.\footnote{For the demonstration purpose, 
we only consider bino-like LSP, i,e, $M_1 < M_2$.} 
with $M_2=1200$ GeV . We further impose the
constraint that the light stop mass ($\mlstop$) is never
heavier than the LSP by more than 30 GeV. For simplicity, the gluino mass and all the other squark and
slepton soft masses are kept fixed at a uniform value, about 100 GeV
above the LSP mass. In principle, these sfermion masses and the gluino mass could 
have been anywhere in between $m_{\wt t_2}$ and $\mlspone$; however, since we are interested in
minimising the mass gap between $m_{\wt t_2}$ and
$\mlstop$ which largely determines the compression factor in
the whole SUSY parameter space, we have kept them at an
intermediate value in order to reduce the number of parameters
to scan. The scan is carried out for two different values of $\tan\beta$, 
namely, 10 and 25 each for two different $\mu$-values (2 TeV and 3 TeV) 
to ascertain their effect on the compression of the
relevant parameter space. 

\begin{figure}[t]
\begin{center}
\includegraphics[trim=50 0 80 0,clip,scale=0.28]{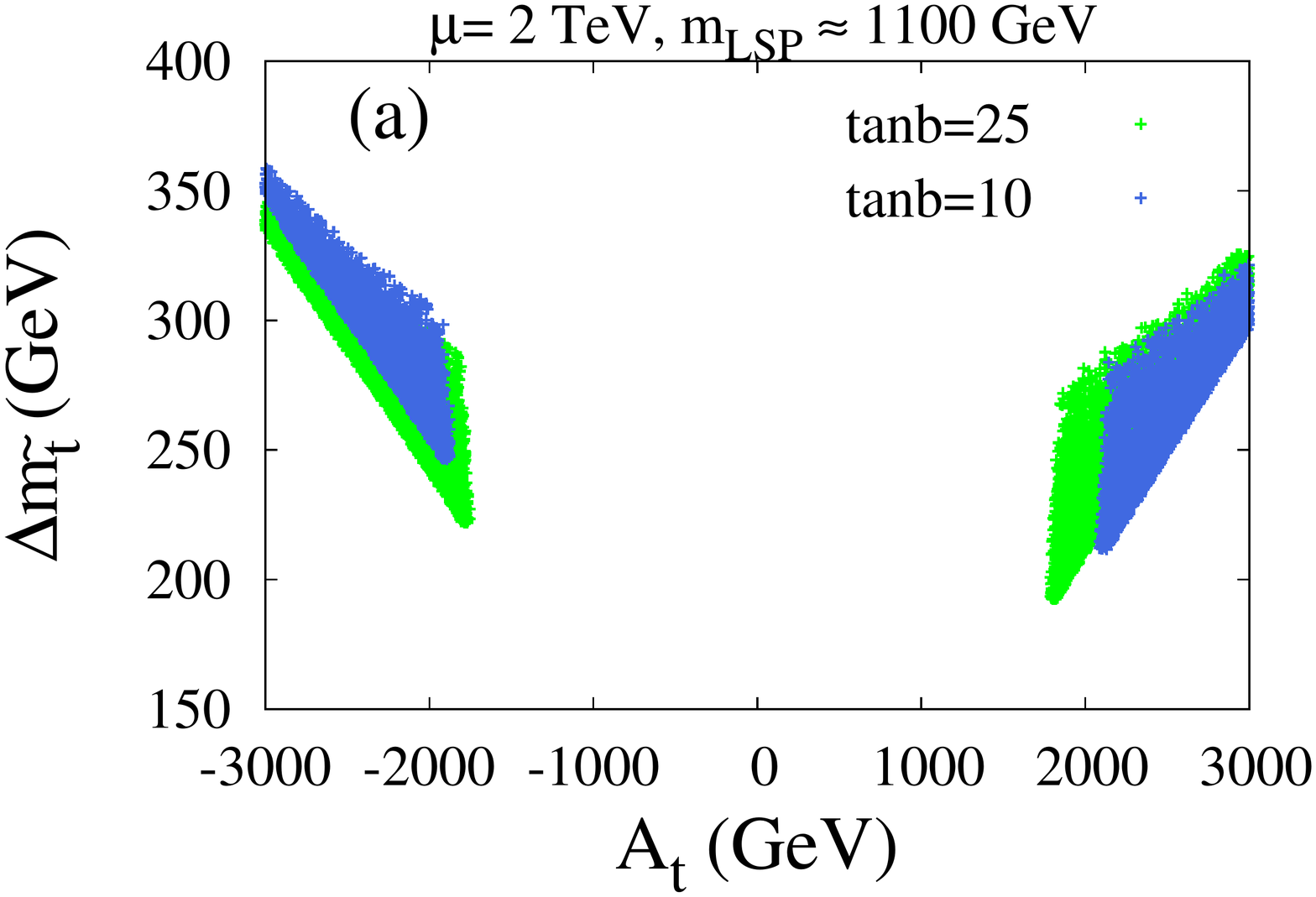}
\includegraphics[trim=50 0 80 0,clip,scale=0.28]{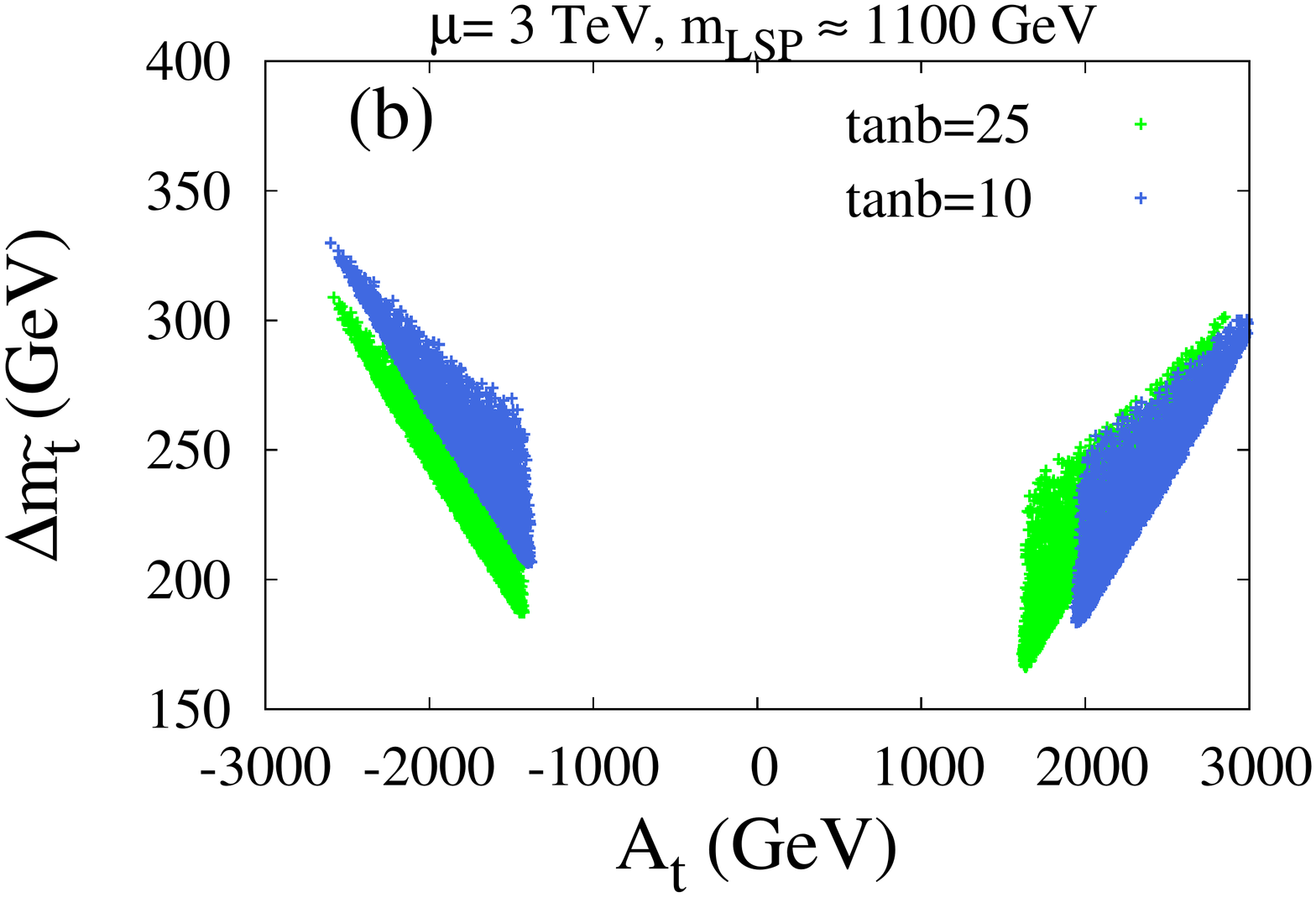}
\includegraphics[trim=50 0 80 0,clip,scale=0.28]{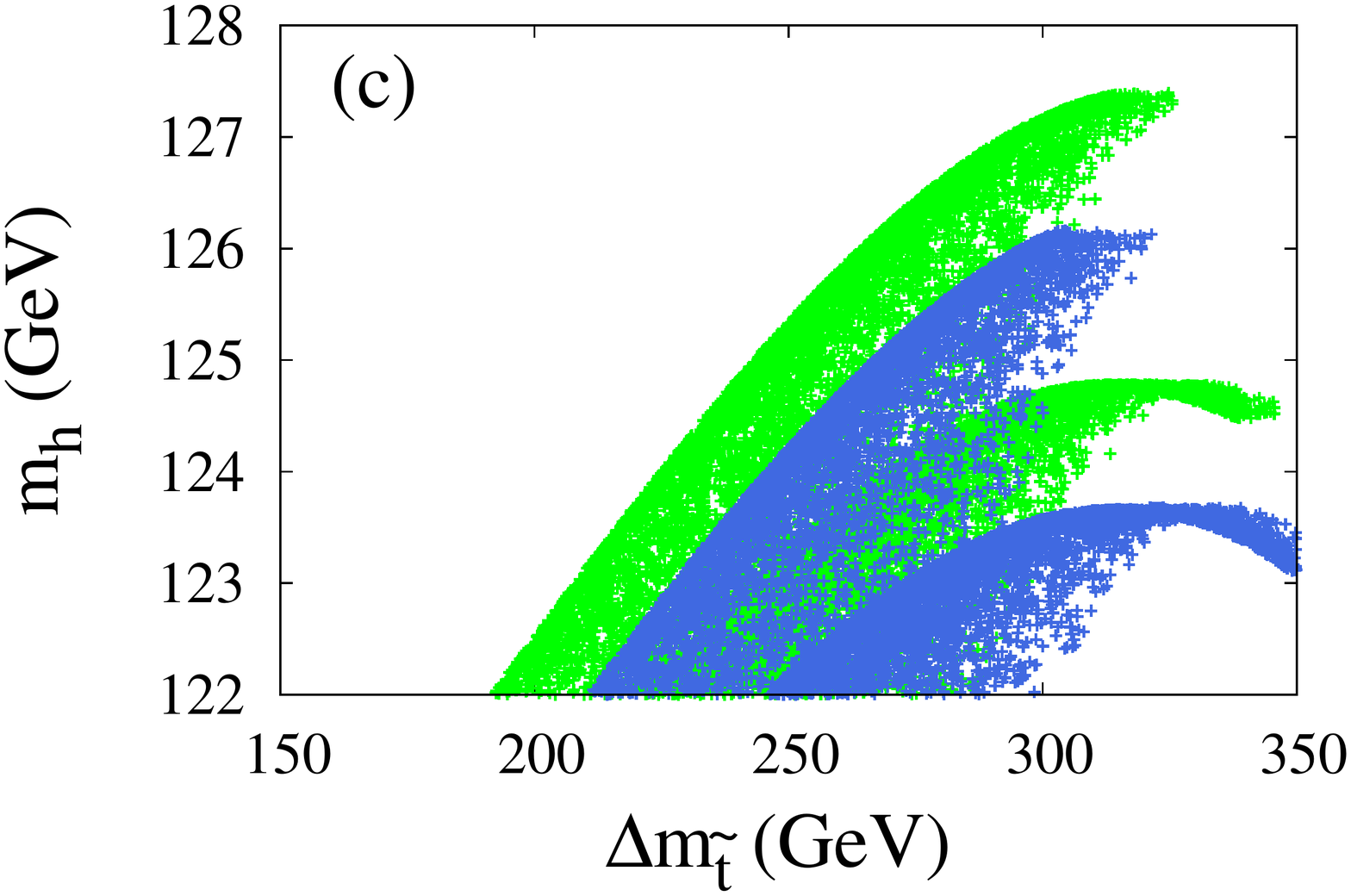}
\includegraphics[trim=50 0 80 0,clip,scale=0.28]{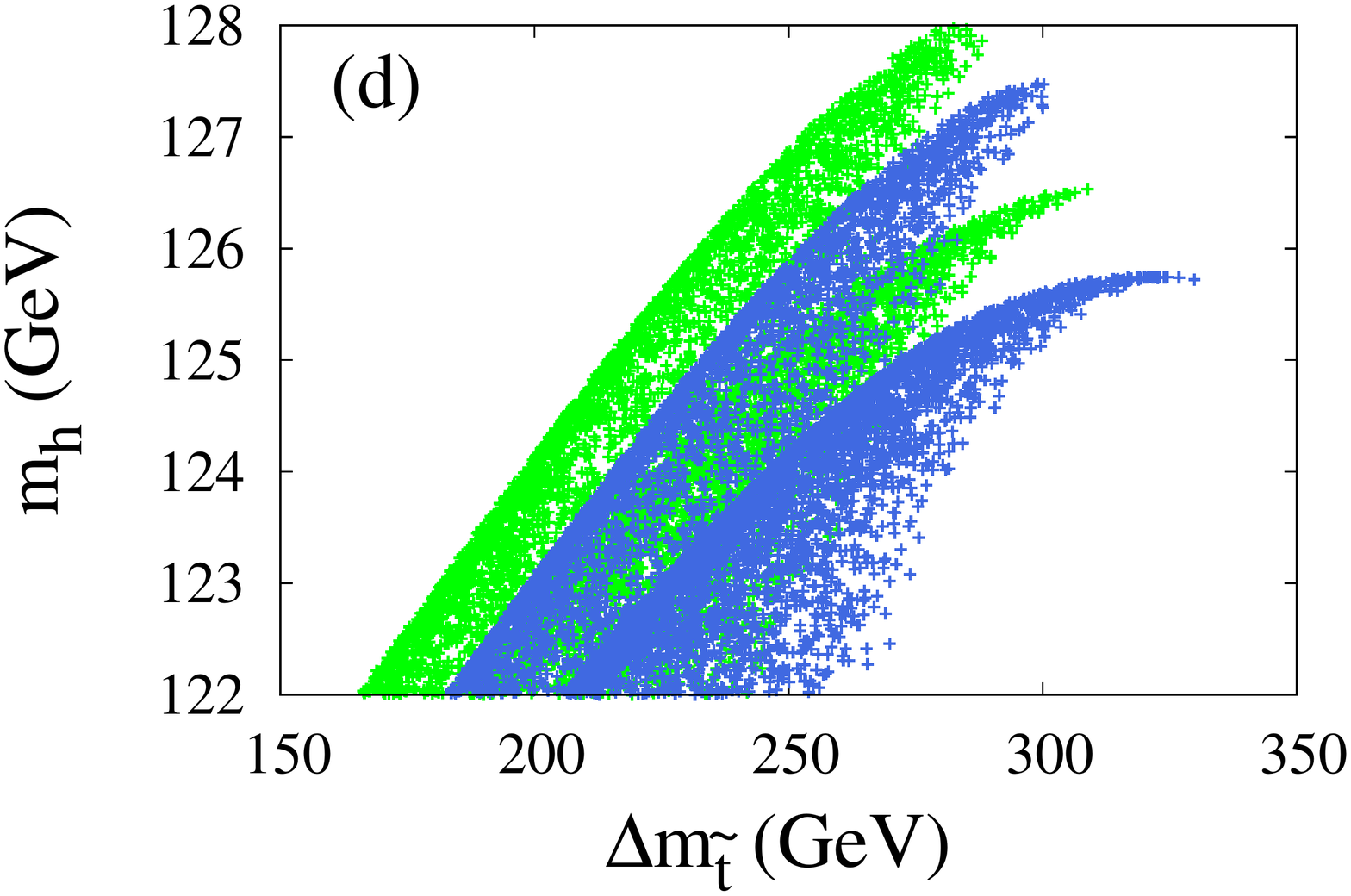}
\includegraphics[trim=50 0 80 0,clip,scale=0.28]{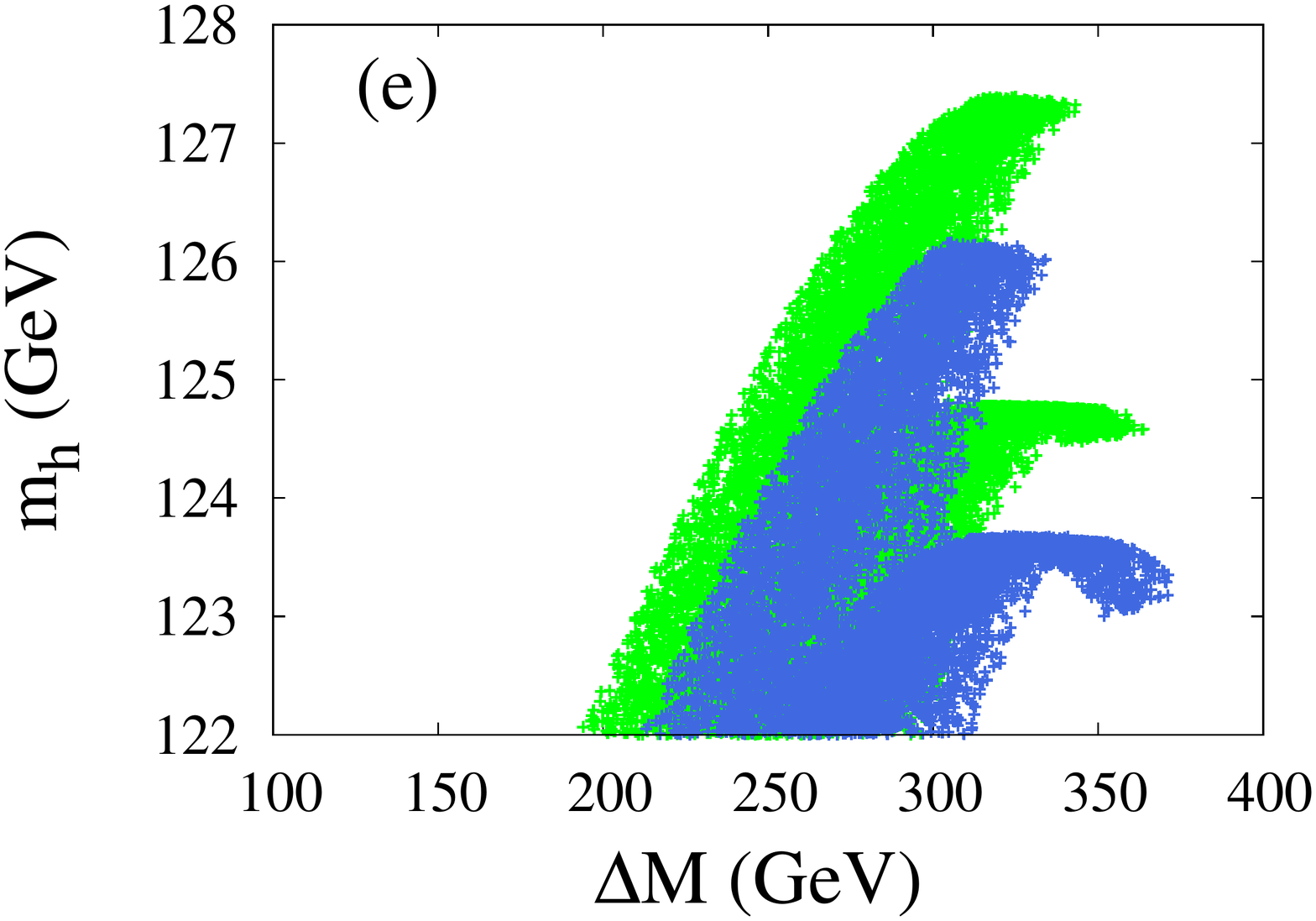}
\includegraphics[trim=50 0 80 0,clip,scale=0.28]{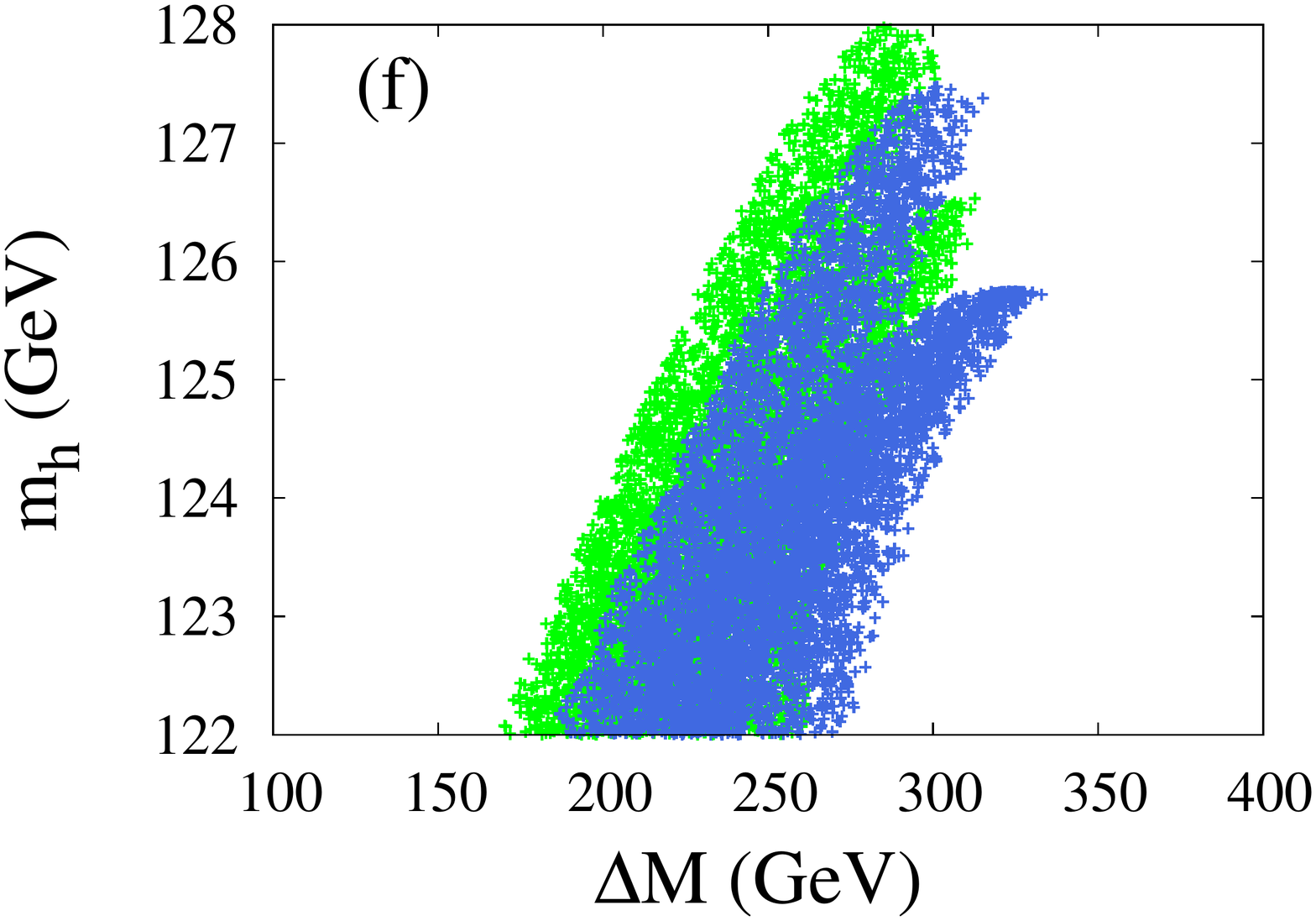}
\caption{The various distributions obtained in our scan are shown. Figs. (a), (c) and (e)  
are obtained with $\mu=2$ TeV while Figs. (b), (d) and (f) show the same set of plots obtained 
with $\mu=3$ TeV. All the points shown in these plots respect the set of constraints 
mentioned in the text. The scan is done for two different $\tan\beta$ values: 25 (green points) 
and 10 (blue points).}
\label{scan_plots_2T_3T}
\end{center}
\end{figure}

Fig.~\ref{scan_plots_2T_3T} showcases the correlation between the 
stop mixing parameters once the
Higgs boson mass constraint is implemented for two different $\mu$
values. As already discussed, the mass difference between the two stop
states, $\Delta m_{\wt t}$, is an important factor in enhancing
the radiative Higgs mass correction. In Figs.~\ref{scan_plots_2T_3T} (a) and (b) 
we show the variation of $\Delta
m_{\wt t}$ with $A_t$ at two different $\tan\beta$
values (Green and Blue points) for $\mu=2$ and 3 TeV respectively. 
As expected, with the increase of $\tan\beta$ smaller $|A_t|$ is allowed from Higgs
mass constraint as a result of increased mixing in stop sector. 
Fig.~\ref{scan_plots_2T_3T} (b) indicates that slightly smaller $|A_t|$ values are
permissible with increase in $\mu$. In a nutshell, the
minimum allowed value of $\Delta m_{\wt t}$ decreases as we
increase $\tan\beta$ or $\mu$ indicating the
possibility of getting more and more compressed spectrum. The minimum
$\Delta m_{\wt t}$ obtained is about 180 GeV with $\mlstop$
close to 1400 GeV and $\mu=2$ TeV whereas with $\mu=4$ TeV this
minimum value reduces to about 100 GeV. 

Figs.~\ref{scan_plots_2T_3T} (c) and (d) show  the distribution of $m_h$ as a  function of
$\Delta m_{\wt t}$. These distributions give a clear idea about
the range  of Higgs mass  we obtaine for  a certain value  of $\Delta
m_{\wt  t}$. Fig.~\ref{scan_plots_2T_3T} (d) shows one can squeeze  $\Delta
m_{\wt t}$ to  about 160 GeV.  However, to  ascertain the whole
sparticle spectrum  mass window, one  needs to look at  the difference
between the  LSP mass and the  heaviest sparticle in  the spectrum. 
Mass gap of the heavier stop/sbottom and
the  LSP  is denoted  as  $\Delta  M$. Figs.~\ref{scan_plots_2T_3T} (e) and (f) show  the
distribution of $m_h$ as a function of $\Delta M$. As evident from the
plots, the minimum $\Delta  m_{\widetilde t}$ is almost similar to 
the minimum $\Delta M$ that is obtained  here indicating that at the 
periphery of this minima, $\mlspone\approx\mlstop$. 

\subsection{Benchmark points}
\label{sec:status:benchmark}
\begin{figure}[ht]
\centering
\frame{\includegraphics[scale=0.45]{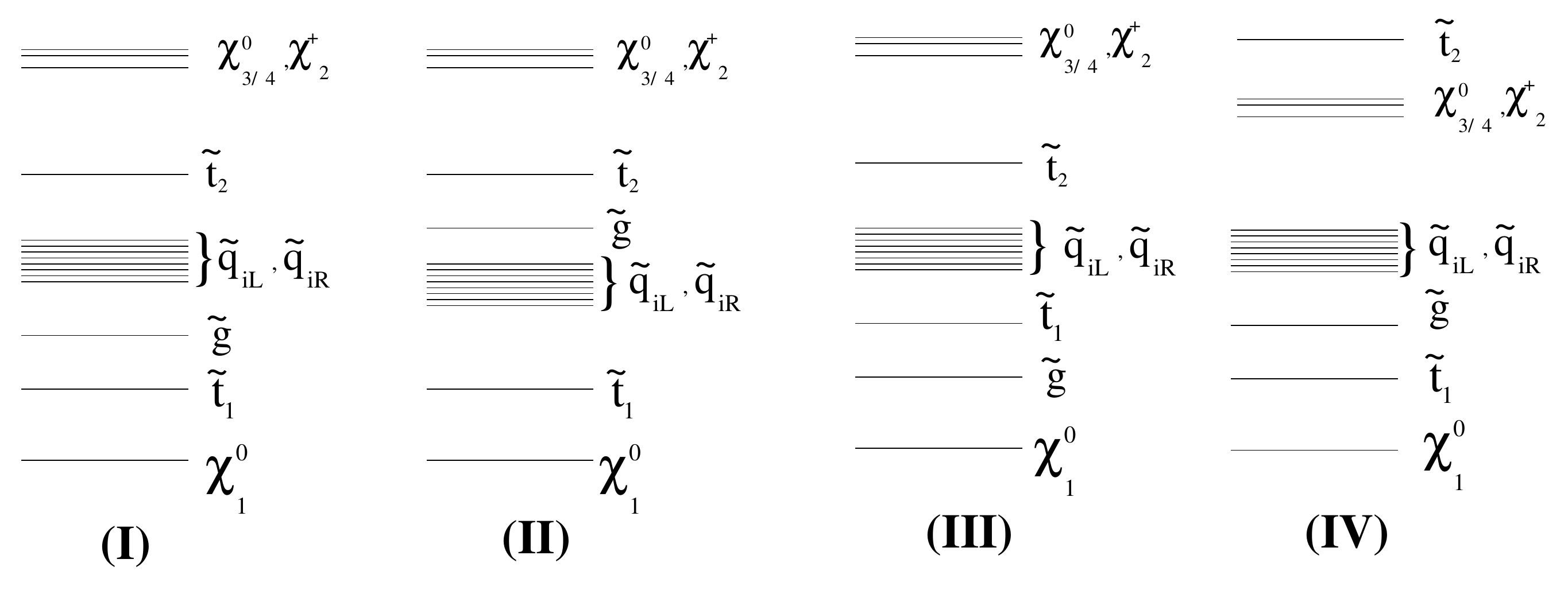}}%
\caption{Different benchmark scenarios considered in our study: Type I ({\tt BP1},{\tt BP3}, {\tt BP5}, {\tt BP10}), 
Type II ({\tt BP4}, {\tt BP7}, {\tt BP8}, {\tt BP9}), Type III ({\tt BP2}) and Type IV ({\tt BP6}).
(In all cases, ${\widetilde{q}_{iL/R}}$=${\widetilde{u}_{iL/R},\widetilde{d}_{iL/R}}$ with ${i=1,2}$. Sleptons, $\chi_2^0$ and $\chi_{1}^{\pm}$ not indicated in the figure,
lie below ${\widetilde{t}_2}$ in all cases. Additionally, the mass gaps shown between different sparticles are not to scale).}
\label{fig:cartoon}
\end{figure}

In choosing the benchmark points for our collider study, we have 
considered a range of LSP masses varying from 840 GeV to 1862 GeV. 
The benchmark choices also take into account a varied mass hierarchy for squarks 
and gluinos, thus allowing different possible decay cascades down to the LSP. 
We also consider situations where the ${\widetilde{g}}$ is the NLSP instead of 
${\widetilde{t}_1}$. An illustrative representation of our choice of benchmark 
points, keeping in mind the different ways the sparticles can be arranged in their 
masses, is presented in  Fig,~\ref{fig:cartoon} where we have classified the benchmarks 
into the four types of representations as shown. 

To study the signal from the above class of spectrum within a compressed SUSY scenario, 
we have chosen ten benchmark points from the allowed parameter space in the model. 
The relevant input parameters, mass spectra and the values of the constraints are summarised in 
Table~\ref{tab:bp_comp}. 

Since having at least one heavy ($\sim$ TeV) stop in the spectrum helps in achieving a Higgs boson mass of 125 GeV, 
it is quite natural to expect more and more compression in the whole SUSY 
spectrum if we keep increasing the LSP mass. In order to showcase this, we have chosen
benchmark points with different LSP masses for different choices of the $\mu$-parameter.
{\tt BP2}, with the lightest LSP mass at 842.4 GeV, gives $\Delta M\sim 300$ GeV while {\tt BP6} has 
the heaviest $\lspone$ at 1861.9 GeV and $\Delta M\sim 184$ GeV. However, note that in {\tt BP6}, 
we are able to even pull down the $\wt\chi_{3/4}$ and the $\wt\chi_2^{\pm}$ masses within 
a 200 GeV mass window from the LSP.  
\begin{center}
\begin{table}[t!]
\scriptsize
\begin{tabular}{||c|c|c|c|c|c|c|c|c|c|c||} \hline\hline
Parameters  & BP1 & BP2 & BP3 & BP4 & BP5 & BP6 & BP7 & BP8 & BP9 & BP10\\ 
\hline\hline 
$M_1$ & 1470.0 & 850.0 & 1107.0 & 1334.5 & 1476.3 & 1890.3 & 1200.0 & 1510.0 & 1105.0 & 1730.0 \\
$M_2$ & 1400.5 & 880.0 & 1200.0 & 1328.6 & 1402.6 & 1971.3 & 1250.0 & 1550.0 & 1150.0 & 1770.0 \\    
$M_3$ & 1312.0 & 780.0 & 1015.0 & 1405.5 & 1387.7 & 1737.1 & 1180.0 & 1420.0 & 1080.0 & 1600.0 \\
$A_t$ & 2200.8 & -1650.0 & 1897.0 & -1535.1 & 1840.8 & 2800.2 & 2050.0 & 2300.0 & 2000.0 & 2720.0 \\
$\mu$ & 2000.0 & 3000.0 & 2000.0 & 3000.0 & 3000.0 & 2000.0 & 2500.0 & 3000.0 & 3000.0 & 2000.0 \\ 
$\tan\beta$ & 20.0 & 20.0 & 25.0 & 23.9 & 24.2 & 16.87 & 18.0 & 20.0 & 20.0 &  35.0 \\
\hline
$m_{\widetilde g}$ & 1429.5 & 861.6 & 1112.8 & 1497.4 & 1500.4 & 1882.0 & 1276.7 & 1534.7 & 1165.6 & 1737.8 \\
$m_{\widetilde q_{L}}$ & 1476.2 & 893.7 & 1159.4 & 1452.3 & 1532.8 & 1912.6 & 1271.6 & 1524.5 & 1129.0 & 1790.0 \\
$m_{\widetilde q_{R}}$ & 1474.3 & 887.4 & 1158.1 & 1451.3 & 1531.9 & 1910.6 & 1270.2 & 1520.8 & 1130.7 & 1794.5 \\
$m_{\widetilde t_1}$ & 1412.3 & 871.7 & 1097.9 & 1330.6 & 1426.1 & 1865.0 & 1192.4 & 1507.6 & 1100.4 & 1711.3 \\
$m_{\widetilde t_2}$ & 1595.9 & 1136.8 & 1300.4 & 1509.0 & 1581.3 & 2045.6 & 1390.5 & 1686.6 & 1308.3 & 1903.2 \\
$m_{\widetilde b_1}$ & 1459.7 & 861.6 & 1137.1 & 1407.4 & 1493.5 & 1966.7 & 1249.6 & 1521.9 & 1130.4 & 1761.3 \\
$m_{\widetilde b_2}$ & 1525.3 & 1044.1 & 1224.7 & 1494.5 & 1570.3 & 2011.6 & 1323.6 & 1619.5 & 1229.4 & 1838.4 \\
$m_{\widetilde\ell_{L}}$ & 1432.7 & 880.9 & 1121.2 & 1400.7 & 1482.7 & 1916.4 & 1221.8 & 1543.9 & 1132.8 & 1745.8 \\
$m_{\widetilde\ell_{R}}$ & 1426.2 & 871.0 & 1114.7 & 1400.7 & 1482.7 & 1907.6 & 1215.4 & 1536.0 & 1121.7 & 1736.9 \\
$m_{\widetilde \tau_1}$ & 1430.3 & 890.3 & 1113.5 & 1353.2 & 1438.0 & 1893.7 & 1220.0 & 1529.1 &  1105.6 & 1725.4 \\
$m_{\widetilde \tau_2}$ & 1483.8 & 1003.3 & 1209.5 & 1446.6 & 1526.0 & 1928.4 & 1289.1 & 1602.2 & 1198.2 & 1803.8 \\
$m_{\widetilde\nu_{L}}$ & 1429.8 & 876.5 & 1117.6 & 1398.6 & 1480.6 & 1914.4 & 1218.3 & 1540.5 & 1128.9 & 1743.1 \\
$m_{\widetilde\chi^0_1}$ & 1406.4 & 842.4 & 1096.3 & 1323.9 & 1417.6 & 1862.0 & 1189.0 & 1496.3 & 1095.4 & 1709.3 \\
$m_{\widetilde\chi^0_2}$ & 1453.9 & 889.1 & 1200.8 & 1342.9 & 1463.6 & 1934.7 & 1256.9 & 1559.0 & 1158.0 & 1764.9 \\
$m_{\widetilde\chi^{\pm}_1}$ & 1406.7 & 889.3 & 1201.0 & 1342.9 & 1417.6 & 1929.1 & 1257.1 & 1559.1 & 1158.2 & 1764.3 \\
\hline
\hline
$m_h$ & 122.6 & 122.0 & 122.2 & 122.5 & 122.8 & 123.9 & 122.0 & 122.4 & 122.1 & 124.6 \\
$\Omega h^2$ & 0.092 & 0.032 & 0.036 & 0.113 & 0.099 & 0.113 & 0.062 & 0.105 & 0.073 & 0.110 \\
$\sigma_{SI}\times 10^{11}$ (pb) & 115.78 & 50.11 & 35.95 & 4.65 & 9.08 & 744.98 & 7.64 & 0.13 & 9.56 & 280.97 \\
$\Delta M $  (GeV) & 189.5 & 294.4 & 204.1 & 185.1 & 163.7 & 183.6 & 201.5 & 190.3 &212.9 & 193.9\\
$\Delta M_i $ (GeV) & 69.8 & 51.3 & 63.1 & 173.5 & 115.2 & 50.6 & 87.7 & 38.4 & 70.2 & 85.2 \\
\hline\hline
\end{tabular}
\caption{Low scale input parameters and the relevant sparticle masses along with the values of the relevant 
constraints for some of the chosen benchmark points satisfying all the collider, DM and low energy constraints 
discussed in this section. All the mass parameters are written in GeV unit. Here, 
$\Delta M_i = m_i-\mlspone$,   
where $i$ represents a gluino or the $1^{st}/2^{nd}$ family squarks (whichever is the heaviest).}
\label{tab:bp_comp}
\end{table} 
\end{center}
A heavier spectrum with $M_1$ or $M_2$ closer to $\mu$ may 
give rise to more compressed spectrum, but they run into trouble with the DM direct
detection constraint. In addition we note that spectra with very heavy squarks and gluino 
would be out of the 13/14 TeV LHC reach with perhaps some hope for the \emph{very high luminosity} 
run. We also take some similar
LSP masses with different squark-gluino mass hierarchies, like in {\tt BP1,
BP5} and {\tt BP3, BP9} to study how the different decay modes and hardness
of jets are affected. It should be noted here that we have focussed on final states with 
zero lepton; one-lepton, two-lepton and three-lepton states have in general highly 
suppressed rates when they arise in cascade decays of coloured sparticles. 
Besides, the leptonic final states often entail backgrounds with harder lepton as well 
as $\met$ spectra, which survive the cuts in a relatively, more abundant manner.
Thus the exact location of the sleptons in our spectra are somewhat inconsequential, 
so far as the multi-jet +{\met} signal is concerned.  
\section{Probing a compressed spectrum at the LHC}
\label{sec:multijet:sigbg}
We explore the possibility of finding such a scenario
with jet(s) + $\met$ final state at the 13 TeV run of the LHC and also
perform a detailed background simulation for the same. 
We consider all possible squark/gluino production channels. 
We must point out that among all the subprocesses contributing to the signal, 
the squark-gluino associated production channel has the largest cross-section 
closely followed by squark pair production cross-section in most of the cases. 
 To study the signal we look at final states with both
mono-jet + $\met$ and multi-jets ($\ge$ 2-jets) + $\met$ in order to
compare the relative statistical significance factors.  

Note that there have been some significant studies  
\cite{Alves:2010za,LeCompte:2011cn,LeCompte:2011fh,Alvarez:2012wf,
  Dreiner:2012gx,Bhattacherjee:2012mz,Dreiner:2012sh,Bhattacherjee:2013wna}
that deal with collider signatures of a compressed spectrum. However, all
these studies consider either squark or gluino pair production and their
subsequent decays into the LSP neutralino. The compression is highlighted through the 
mass gap between the squark/gluino and the LSP being small, begging the explanation that
the final state jets in such cases are too soft to be detected at the colliders.  
In order to observe any signal, one then has to rely on the ISR-FSR jets and/or photons. 
While such an observation may shed light on a somewhat fine tuned compression in the SUSY
spectrum, one cannot fathom that no other SUSY particle will be in similar mass ranges.
We believe that we have already highlighted that an equally probable spectrum, where 
almost all SUSY particles are squeezed within a relatively small mass gap between the LSP 
and the heaviest coloured sparticle, meets the strictest of experimental constraints there is to
offer. Such a scenario, therefore, presents a situation where one can envisage additional 
contributions to the final states in consideration through production of the closely lying coloured
sparticles. Through this work we try to show how this could lead to modifications in the signal 
topologies and what optimisations in kinematic selections may be required to study 
such a compressed SUSY signal at the LHC.
\subsection{Analysis setup and simulation details}
We consider all possible production channels of the coloured sparticles, \emph{i.e.} 
\begin{align*}
 pp \to {\widetilde{q}_i \, \widetilde{q}_j}, \,\,\, {\widetilde{q}_i \, \widetilde{{q}_j}^*}, \,\,\, {\widetilde{g} \, \widetilde{g}}, \,\,\, {\widetilde{q}_i \, \widetilde{g}}, \,\,\, {\widetilde{q}^*_i \, \widetilde{g}}
\end{align*}
where the respective sparticles would cascade down to the LSP, giving a multi-particle final 
state comprising of leptons and quarks along with $\met$ associated with the invisible LSP. It turns 
out that for the compressed spectrum, the jets and charged leptons originating from cascade decays 
are expected to be quite soft.  Therefore it becomes quite likely that events observed from such 
productions could be observed through jets originating from initial-state radiation (ISR). 
As a trigger threshold for such jets would naturally include situations where the jets may 
actually be coming from hard partons produced in association with the pair of SUSY particles at the 
parton-level. Hence one necessarily requires to produce hard jet(s) at the parton
level along with the coloured sparticles and match the events with the ISR jet events. 
We perform a parton level event generation simulation 
using {\tt MadGraph5}(v2.2.3) \cite{Alwall:2014hca,Alwall:2011uj}. 
For our analysis we have chosen {\tt CTEQ6L} \cite{Pumplin:2002vw} as the parton distribution function (PDF). 
The factorisation scale is set following the default option of {\tt MadGraph5}.
The generated events are passed through {\tt PYTHIA}(v6.4) \cite{Sjostrand:2006za} 
to simulate showering and hadronisation effects, including fragmentation. 
The matching between shower jets and jets produced at the parton level is done using {\tt MLM} 
matching \cite{Mangano:2006rw,Hoche:2006ph} based on $shower$-$k_{T}$ algorithm with $p_T$-ordered showers. 
The matching scale, defined as {\tt QCUT}, differs for the signal where heavy SUSY particles are 
produced in association with jets when compared to the scale chosen for the SM background. 
Typical choice of this scale is set between $\simeq 20-30$ GeV for the SM backgrounds, 
and $\simeq 100-120$ GeV for the MSSM processes after careful investigation of the matching plots 
generated for different {\tt QCUT} values.
Then we pass the events through {\tt Delphes}(v3.2.0) 
\cite{deFavereau:2013fsa,Selvaggi:2014mya,Mertens:2015kba} for jet formation, using 
$anti$-$k_T$ jet clustering algorithm \cite{Cacciari:2008gp} $($via {\tt FastJet} 
\cite{Cacciari:2011ma}$)$, and detector simulation with default ATLAS selection cuts. 

As the signal under consideration is either mono-jet $+\met$ or multi-jet $+\met$, we need to identify 
the dominant SM subprocesses that can contribute to the above. For hadronic final states, the 
most dominant contribution comes from the pure QCD processes such as multi-jet production where
$\met$ comes either from the jets fragmenting into neutrinos or simply from mismeasurement of the 
jet energy.  Other significantly large contributions can come from $W$ + jets where the $W$ decays 
leptonically and the charged lepton is missed, $Z$ + jets where the $Z$ decays to neutrinos and 
$t\bar{t}$ production. 
Additional modes that may also contribute include $t$ + jets and $VV$ + jets where $V=W^\pm,Z$.
For reasons already stated in section \ref{sec:status:benchmark}, a lepton veto in the final state 
helps to suppress quite a few of the above backgrounds. 
The matching scheme has been also included for the SM background wherever necessary.
\subsubsection*{Primary selection criteria}
\label{Primary selection criteria}
To identify the charged leptons ($e,\mu$), photon ($\gamma$) and jets,  we put the following basic selection criteria ({\bf C0}) on the final state particles for both signal and background: 
\begin{itemize}
\item Leptons ($\ell = e,\mu$) are selected with $p_T^{\ell} > 10$  GeV, $|\eta^{e}| < 2.47$ and $|\eta^{\mu}| < 2.40$, excluding the transitional pseudorapidity region between the barrel and end cap of the calorimeter $1.37 < |\eta^{\ell}| < 1.52$.
\item Photons are identified with $p_T^{\gamma} > 10$ GeV and $|\eta^{\gamma}| <  2.47$ excluding the same transition window as before.
\item  We demand hard jets having $p_T^{j} > 40$ GeV within $|\eta^j| < 2.5$.
\item All reconstructed jets are required to have an azimuthal separation with $\vec{\slashed{E}}_T$ 
given by $\Delta\phi({\rm jet},\vec{\slashed{E}}_T)  > 0.2$.
\end{itemize}
\begin{figure}[t]
\begin{center}
\includegraphics[trim=0 0 50 0,clip,scale=0.28,width=0.47\linewidth,height=0.43\linewidth]{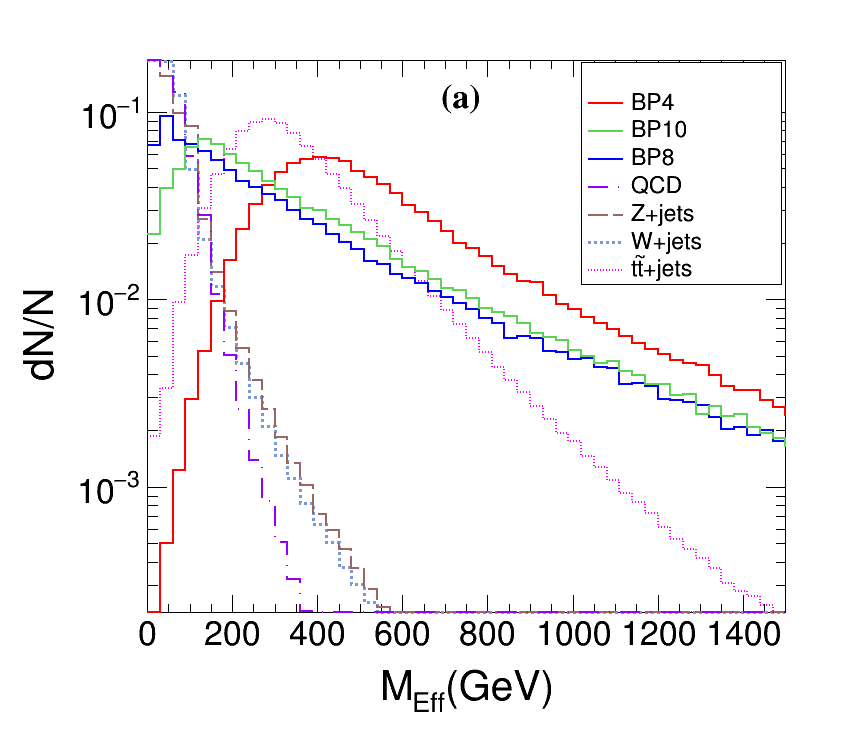}
\includegraphics[trim= 0 0 22 0,clip,scale=0.28,width=0.48\linewidth,height=0.43\linewidth]{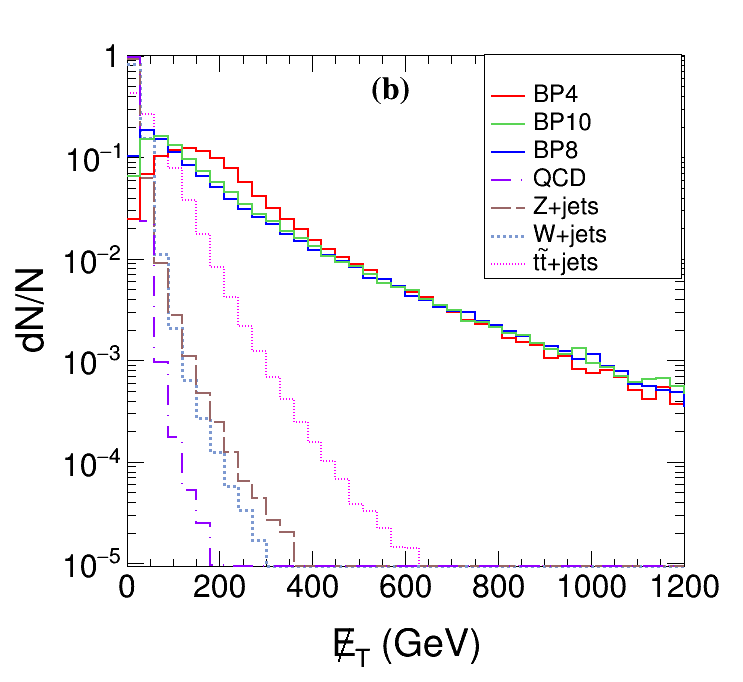}\
\includegraphics[trim=0 0 40 0,clip,scale=0.28,width=0.47\linewidth,height=0.44\linewidth]{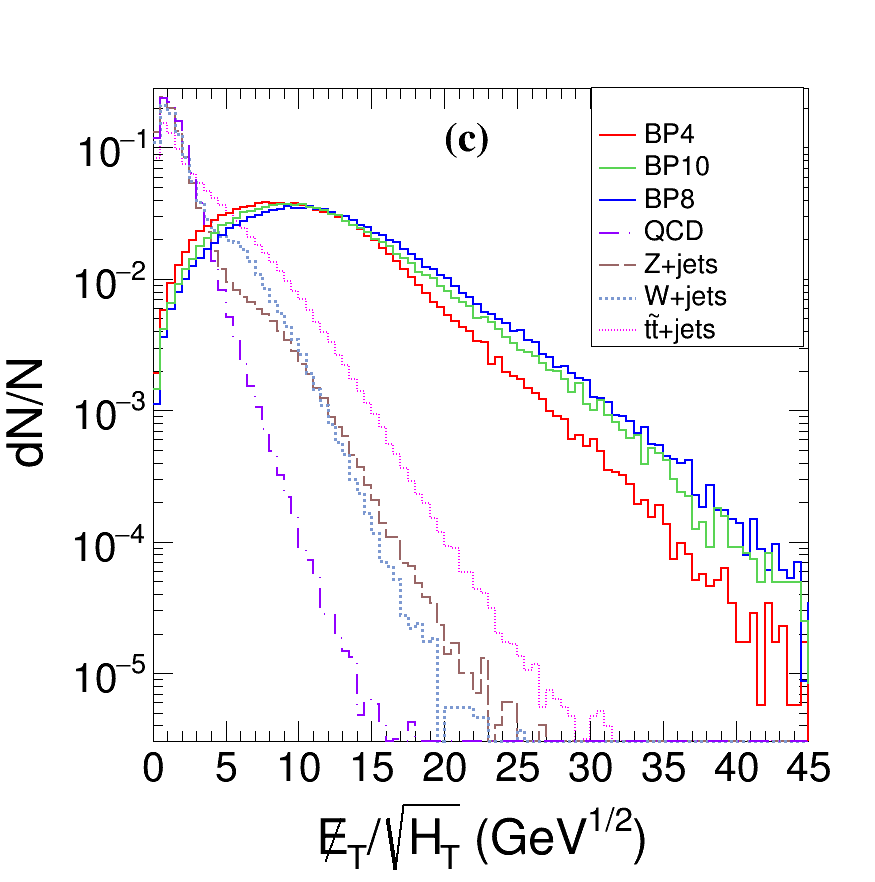}
\includegraphics[trim=0 0 45 0,clip,scale=0.28,width=0.48\linewidth,height=0.43\linewidth]{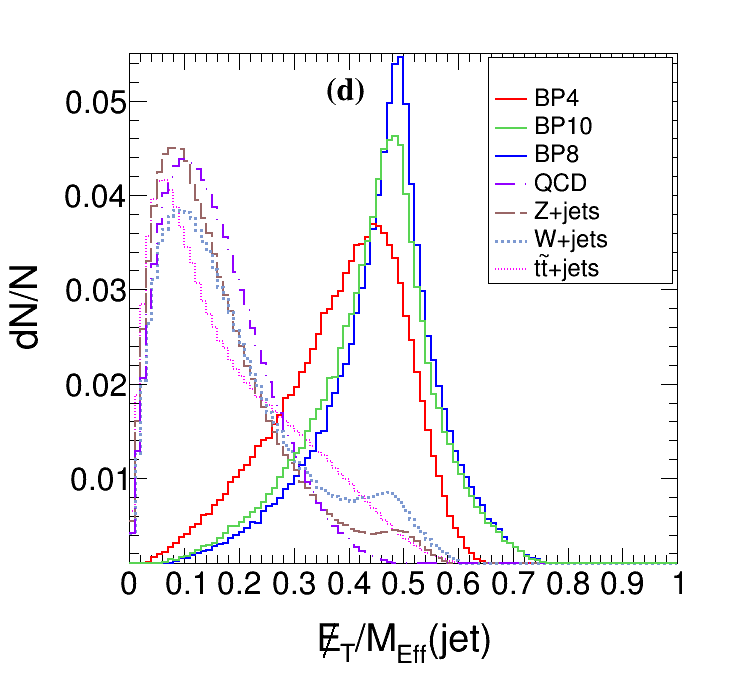}\
\caption{Normalised differential distributions of a few relevant kinematic variables for our analysis of 
compressed spectra after imposing the event selection cuts {\bf C0}. For illustration, 
signals {\tt BP4, BP8} and {\tt BP10} are compared to the SM Backgrounds. See the draft for the description of $M_{Eff}(jet)$.}%
\label{fig:kin1}
\end{center}
\end{figure}
Once the primary selection criteria are set for the signal and background events, we now 
need to identify specific kinematic characteristics that would differentiate the SUSY events from that 
of the SM background. To highlight the differences, we choose for illustration a few benchmark points, namely {\tt BP4, BP8} and {\tt BP10}. 
In Fig.~\ref{fig:kin1} we show the normalised distributions of some relevant kinematic variables where 
one can expect significant differences between the signal events of SUSY and the SM, after imposing 
the above selection criteria {\bf C0}. Note that events with all jet multiplicities have been included in these plots. 
As the SUSY signal arises from production of heavy 
coloured sparticles and is expected to carry large missing energy due to the invisible heavy LSP's in 
the final state, the effective mass ($M_{Eff}$) and $\met$ are expected to help in differentiating the 
SUSY events from SM.
In Fig.~\ref{fig:kin1}(a) we present the effective mass ($M_{Eff}$) distribution for these channels where 
\begin{align*}
M_{Eff} = \sum_i |\vec{p}_{T_{i}}| + \met 
\end{align*}
and $i$ runs over all the states present in the event including the reconstructed jets. This global variable, 
without utilising any topology information, can be extremely efficacious from the understanding that, 
contrary to most of the SM background processes, production of heavy SUSY particles require 
significantly larger parton level center-of-mass (CM) energy. Thus one expects a larger 
$M_{Eff}$ for all benchmark scenarios as shown in  Fig.~\ref{fig:kin1}(a).  
In Fig.~\ref{fig:kin1}(b) we show the expected missing transverse energy distributions for the 
SUSY signal and SM background. Quite clearly, the distributions in both the above variables 
seem to peak at lower values for the SM background (except $t\bar{t}$+jets) when compared to the 
SUSY signal. Note that we have plotted the normalised distributions which gives a qualitative idea 
on the additional cuts required on these variables, rather than a quantitative one. 

In addition we find two more kinematic variables of interest used by the ATLAS 
Collaboration \cite{Aad:2014wea,Aad:2015zva}, \emph{viz.}  $\met/\sqrt{H_T}$ and $\met/M_{Eff}(jet)$, which 
show clear difference between signal and background. These are shown in Fig.~\ref{fig:kin1}(c) 
and Fig.~\ref{fig:kin1}(d) respectively. Here, $H_T$ represents the scalar sum of all isolated jet 
$p_T$'s while $M_{Eff}(jet)$ is defined to be constructed out of the first two leading jets and $\met$:
\begin{align*}
M_{Eff}(jet) =  p_{T}^{j_1} + p_{T}^{j_2} + \met 
\end{align*}
These plots also show some distinct characteristic distributions for signals. We thus find that 
appropriate cuts on the above variables, shown in Fig.~\ref{fig:kin1}, would serve to optimise the signal
vis-a-vis the SM background. We now proceed to analyse the multi-jets + $\met$  and mono-jet + $\met$ signals in the next
section.
%
\subsection{Multi-jets $+\met$}
\label{sec:multijet}
As discussed earlier,  a compressed SUSY spectra such as ours can lead to high multiplicity of jets in the final state. We observe that significant signal events are found when the jet-multiplicity ($n_j$) is 
at least two ($n_j \geq 2$) after selecting events using {\bf C0}.  This multi-jet scenario is dependent 
on the hardness of the selected jets and therefore one requires optimised event selection criteria 
to see how it stands against the SM background. We list below the different cuts which help 
us in achieving an improved signal to background ratio:   %
\begin{itemize}
\item {\bf C1}: Since we are only considering squark and gluino production channels, no hard lepton or photon are expected in the final state. We, therefore, select final states with two or more jets, vetoing any qualified lepton or photon in such events.
\end{itemize}
\begin{table}[t]
\small
\begin{center}
\footnotesize
\begin{tabular}{|c|c|c|c|c|c|c|c|c|} 
\hline 
\multicolumn{2}{|c|}{Signal} & 
\multicolumn{7}{|c|}{Effective cross-section after the cuts (in fb)} \\
\hline
Benchmark & Production               
& {\bf C0 $ + $ C1} & {\bf C2} & {\bf C3} & {\bf C4} &  {\bf C5}  & {\bf C6} & {\bf C7} \\
 Points   &  cross-section(fb)  & &    
        &      &    &    &    &     \\
\hline  
BP1 & 155.56 & 87.38 & 24.32 & 23.34 & 11.49 & 11.29 & 8.28 & 8.22 \\ 
BP2 & 4202.42 & 1877.45 & 588.58 & 564.81 & 260.89 & 255.29 & 176.81 & 175.21 \\
BP3 & 835.49  & 414.61 & 126.64 & 121.58 & 58.32 & 57.12 & 40.96 & 40.66  \\
BP4 & 126.93 & 118.79 & 62.85 & 59.72 & 20.74 & 19.84 & 9.99 & 9.93\\
BP5 & 93.77 & 81.83 & 41.58 & 39.57 & 13.64 & 13.17& 7.18 & 7.13 \\
BP6 & 29.66 & 14.39 & 5.49 & 5.30 & 2.76 & 2.71 & 2.03 & 2.01 \\ 
BP7 & 364.38 & 248.29 & 82.04 & 77.54 & 32.81 & 31.99 & 20.16 & 20.04\\
BP8 & 95.58 & 40.62 & 12.86 & 12.45 & 6.34 & 6.24 & 4.72 & 4.68  \\
BP9 & 731.08 & 453.91 & 117.84 & 112.37 & 55.17 & 53.86 & 35.92 & 35.62   \\
BP10 & 29.60 & 19.21 & 5.20 & 4.99 & 2.37 & 2.33 & 1.65 & 1.64 \\
\hline 
\end{tabular}
\caption{The cut-flow table for the (multi-jet $+\met$) final state, showing the change in signal 
cross-sections for the ten different benchmark points. The cuts ({\bf C0 -- C7}) are defined in the text in 
Sec. \ref {sec:multijet}.}
\label{signals1NLO} 
\end{center}
\end{table} 
%
The multi-jet signal is defined for events that satisfy {\bf C0 + C1}. Note that for a compressed SUSY 
spectrum, the jet multiplicity would start falling when more hard jets are selected in the final state. 
An optimised choice in our case is to have only a few very hard jets with the following 
requirements on their transverse momenta: 
\begin{itemize}
\item {\bf C2}: The hardest jet should have $p_T(j_1) > 130$ GeV and the next hardest jet $p_T(j_2) > 80$ GeV.
\end{itemize}
%
\begin{table}[h]
\small
\begin{center}
\footnotesize
\begin{tabular}{|c|c|c|c|c|c|c|c|c|}    
\hline
\multicolumn{2}{|c|}{SM Backgrounds} &
\multicolumn{7}{|c|}{Effective cross-section after the cuts (in pb)} \\
\hline
Channels & Production                    
&  {\bf C0 $ + $ C1} & {\bf C2} & {\bf C3} & {\bf C4} &  {\bf C5}  & {\bf C6} & {\bf C7} \\
       & (in pb)  & &      
        &      &    &    &    &     \\
\hline 
$\rm t \bar{t}$ +${\leq}$ 2 jets & 722.94 & 542.67 & 167.2 & 141.63 & 15.54 & 2.47 & 0.16 & 0.151\\
t +${\leq}$ 3 jets & 330.57 & 227.0 & 36.23 & 29.84 & 1.09 & 0.123 & 0.01 & 0.009   \\
QCD(${\leq}$4 jets)& 2E+08 & 1.8E+07 & 312747 & 251865 & 2765.52 & ${\sim}$0 & ${\sim}$0 & ${\sim}$0 \\   
Z +${\leq}$ 4 jets & 57088 & 6660.86 & 325.92 & 265.45 & 13.39 & 2.10 & 0.666  & 0.666  \\
W + ${\leq}$4 jets &  197271 & 14206.3  & 896.76 & 734.47 & 36.93 & 3.98 & 0.485 & 0.485    \\
WZ + ${\leq}$2 jets & 53.8 & 24.44 & 5.74 & 4.81 & 0.67 & 0.16 & 0.037 & 0.036\\
ZZ + ${\leq}$2 jets & 13.69 & 5.77 & 0.79 & 0.66 & 0.069 & 0.019 & 0.00549 & 0.00548 \\

\hline 
Total & \multicolumn{7}{|c|}{} &  \\
background & \multicolumn{7}{|c|}{} & 1.352 \\
\hline 
\end{tabular}
\caption{
The cut-flow table for the (multi-jet $+\met$) final state, showing the change in cross-sections for 
the different subprocesses contributing to the SM background. The cuts ({\bf C0 -- C7}) are defined in 
the text in Sec. \ref {sec:multijet}.}
\label{backgrounds1} 
\end{center}
\end{table}
We find that the above requirement does not affect the signal significantly while giving appreciable 
suppression to the SM background (see Table~\ref{signals1NLO} and Table~\ref{backgrounds1}).
\begin{itemize}
\item {\bf C3}: We demand larger azimuthal separations between the leading two jets and 
$\vec{\slashed{E}}_T$  {\it i.e.} $\Delta\phi({\rm jet},\vec{\slashed{E}}_T) > 0.4$.  This requirement is  necessary to reduce the chance of contamination in the $\met$ coming from missing parts of 
these hard jets. 
\end{itemize}
We note that the above set of requirements ({\bf C0 -- C3}) not only helps in refining the signal 
against the SM background but also helps us in determining more precise quantitive cuts
on the kinematic variables shown in Fig.~\ref{fig:kin1} to improve the signal significance 
(see Table~\ref{signals1NLO} and Table~\ref{backgrounds1}). Naively, Fig.~\ref{fig:kin1}(d) 
would suggest that an appropriate cut on $\met/M_{Eff}(jet)$ itself can help us completely eliminate 
the background. However, on close inspection, we find that the tail of the large QCD background still 
survives this cut. We, therefore, find a more optimised cut flow to improve the signal significance as 
shown below.       
\begin{itemize}
\item {\bf C4}: We demand $M_{Eff} > 800$ GeV. This turns out to be quite crucial in 
significantly suppressing almost all contributions for the SM background while moderately 
affecting the signal events. 
\item {\bf C5}: We demand $\met >  160$ GeV which helps in completely eliminating the remnant 
QCD multi-jet background while suppressing all the other SM background channels. Note that 
this cut hardly affects the signal for any of the benchmark points. 
\item {\bf C6}: Larger missing energy and softer jets in our scenario  results in a larger $\met/\sqrt{H_T}$ 
ratio compared to the SM background channels. We find that with $\met/\sqrt{H_T}  > 15$ GeV$^{1/2}$ 
the signal significance can be improved further.
\item {\bf C7}: The ratio $\met/M_{Eff}(jet)$ is shown to peak at smaller values for the SM background and
therefore we demand $\met/M_{Eff}(jet) >0.35$ which further improves our signal significance.
\end{itemize}

\begin{table}[t]
\small
\begin{center}
\begin{tabular}{|c|c||c|c|c||c|c|c|}
\cline{3-7}
\multicolumn{1}{c}{} &
\multicolumn{1}{c||}{}&
\multicolumn{3}{|c||}{{Statistical significance}} & \multicolumn{2}{|c|}{Required luminosity} \\
\multicolumn{1}{c}{} &
\multicolumn{1}{c||}{}&
\multicolumn{3}{|c||}{{(${\mathcal S}$)}} & \multicolumn{2}{|c|}{(in $fb^{-1}$)} \\
\hline
Signal & $ \rm m_{\widetilde{\chi}_1^0}$({\rm GeV}) & 
${\mathcal L}=100 $ & 
${\mathcal L}=500 $  & ${\mathcal L}=1000 $  & ${\mathcal S}=3{\sigma}$ & ${\mathcal S}=5{\sigma}$  \\ 
\cline{1-7}
BP1 & 1406.4 
& 2.23 
& 4.99 & 7.06  & 180.98 & 502.72 \\ 
BP2 & 842.4 
&46.67  &
104.37 & 147.60& 0.41 & 1.15  \\
BP3 & 1096.3  
&11.00  & 
 24.61& 34.80 & 7.44 & 20.66  \\
BP4 & 1323.9  
&2.70 
& 6.03 & 8.53 &  123.46 & 342.94\\ 
BP5 & 1417.6 
&1.94  
&4.33 &6.13 &239.13 & 664.26 \\ 
BP6 & 1862.0 
&0.55  
&1.22 & 1.73  &  2975.21  & 8264.46  \\ 
BP7 & 1189.0 
&5.44 &
12.16 & 17.19 & 30.41 & 84.48\\
BP8 & 1496.3
 &1.27 & 
2.84 & 4.02 &  558.00 & 1550.00\\
BP9 & 1095.4 
&9.65&
21.57 & 30.50 & 9.66 & 26.85\\ 
BP10 & 1709.3 &
0.45 &
1.00 & 1.41 & 4444.44 & 12345.68 \\
 \hline
\end{tabular}
\caption{Statistical significance of the signal for different benchmark points in the 
multi-jet $+\met$ analysis at 13 TeV LHC.  The significance is estimated for three values of integrated luminosity (${\mathcal L}=100, 500$ and $1000$ fb$^{-1}$). We also estimate the required integrated 
luminosity to achieve a $3\sigma$ and $5\sigma$ excess for each benchmark point at LHC 
with $\sqrt{s}=13$ TeV.}
\label{multi-significanceNLO}
\end{center}
\end{table}
We present the numerical results for the ten aforementioned benchmark points and the
SM background to the multi-jet $+\met$ signal at LHC with $\sqrt{s}=13$ TeV.   
In Table~\ref{signals1NLO} we summarise the effects of the cuts ({\bf C0 -- C7}) on signal cross-sections. 
It is worth pointing out here that we have used next-to-leading order (NLO) cross-section for the 
production rates of squarks and gluinos in our signal analysis by multiplying the leading-order 
cross-sections of {\tt MadGraph5} by NLO $K$-factors obtained using Prospino 2.1 \cite{Beenakker:1996ed,Plehn:2004rp,Spira:2002rd,Beenakker:1997ut,Beenakker:1996ch}.  
The cut-flow table for the same set of cuts is shown for the SM background processes in 
Table~\ref{backgrounds1}.  Note that we have also used the NLO cross-section for SM 
background processes provided in {\tt MadGraph5} \cite{Alwall:2014hca}. 
It is quite clear to see from Tables~\ref{signals1NLO} and \ref{backgrounds1} that our choice of cuts 
is quite efficient in suppressing a seemingly huge SM background such that the signal 
may be observed at the LHC. The  
statistical significance {(${\mathcal S}$)} of the observed signal ($s$) over the total SM background 
($b$) has been calculated using
\begin{equation}
{\mathcal S} = \sqrt{2 \times \left[ (s+b){\rm ln}(1+\frac{s}{b})-s\right]}. 
\label{sig:eqn}
\end{equation}
We show the significance of the signal for different benchmark points in the 
multi-jet $+\met$ channel in Table~\ref{multi-significanceNLO}. We choose three different values for 
the integrated luminosity (${\mathcal L}=100, 500$ and $1000$ fb$^{-1}$) to highlight the significance 
for the benchmark points. We find that the signal corresponding to {\tt BP6} and {\tt BP10} are 
practically scenarios within a compressed SUSY spectrum which would be very hard to observe 
in the multi-jet $+\met$ channel. In fact an integrated luminosity of over $3000$ fb$^{-1}$ would be 
required for any hope of observing a notable excess for the SUSY spectrum given 
by the above benchmark points. This is however understandable as the corresponding spectra is 
very heavy leading to smaller production rates compared to the other benchmark points. The 
lightest spectrum amongst all the benchmark points, \emph{viz.} {\tt BP2} is the most robust of all
and should be observable at the present run of LHC with luminosity as low as 1 fb$^{-1}$.  
The rest of the benchmark points too lead to $3 \sigma$ and $5 \sigma$ excess over the 
SM backgrounds with relatively nominal to slightly higher integrated luminosities as shown in the 
last two columns of Table~\ref{multi-significanceNLO}.
\subsection{Mono-jet $+\met$}
\label{sec:monojet}
The mono-jet +$\met$ signal is considered as a favourable channel to probe a compressed spectrum 
at the LHC \cite{Alves:2010za,Alvarez:2012wf,Dreiner:2012gx,Bhattacherjee:2012mz,Dreiner:2012sh,Bhattacherjee:2013wna,
Martin:2008aw,Belanger:2012mk,Cohen:2013xda,Mukhopadhyay:2014dsa,Han:2014xoa}. 
Therefore it is quite logical to explore how the mono-jet final state in our scenario 
stands against the SM background. Both ATLAS and CMS have investigated mono-jet signals in the context of 
compressed SUSY spectra \cite{Chatrchyan:2011nd,Aad:2011xw,Aad:2012fqa,Khachatryan:2014rra,Aad:2015zva}. 
Note that, these analyses consider only such scenarios where the compression 
is between the NLSP and LSP, and the signal arises through the NLSP pair production and 
its subsequent decay. Since we consider almost the entire SUSY spectrum to be compressed, 
all SUSY processes (dominated, of course, by coloured sparticle production channels) are of interest to us. 
Thus our analysis requires revisiting the standard cuts suggested in the literature. As in the case of 
multi-jet $+\met$ final state, we demand a leptonically quiet mono-jet final state (after {\bf C0}) where:
\begin{itemize}
\item {\bf D1}: Events are selected having at least one hard jet in the final state with no charged lepton 
or photon.
\end{itemize}
Since mono-jet searches rely on hard ISR jet, the leading jet is required to be considerably hard with
large transverse momentum and well separated  from the direction of $\vec{\slashed{E}}_T$:
\begin{itemize}
\item {\bf D2}: The leading jet has $p_T (j_1)> 130$ GeV (as before) with a significantly larger 
azimuthal separation with $\vec{\slashed{E}}_T$ given by ${\Delta\phi(j_1,\vec{\slashed{E}}_T) > 1.0 }$.
\item {\bf D3}: In order to accommodate a hard jet coming from ISR, but not rule out cases 
with another jet arising due to its fragmentation, we demand the second hardest jet to  
have $p_T(j_2) < 80$ GeV, but with ${\Delta\phi(j_2,\vec{\slashed{E}}_T) > 1.0 }$.
\end{itemize}
We thus define our mono-jet + $\met$ 
signal for events which satisfy cuts ({\bf C0, D1 -- D3}). 
Moreover, for events where the leading jet is hard enough, a sizeable $\met$ is seen,  
which also serves as an useful discriminator for the SUSY signal from the SM background. 
\begin{table}[t]
\small
\begin{center}
\footnotesize
\begin{tabular}{|c|c|c|c|c|c|c|c|} 
\hline 
\multicolumn{2}{|c|}{Signal} & 
\multicolumn{5}{|c|}{Effective cross-section after the cuts (in fb)} \\
\hline 
Benchmark & Production               
&{\bf C0 $ + $ D1} & {\bf D2} & {\bf D3} & {\bf D4} &  {\bf D5}    \\
 Points   &  cross-section(fb)  & &    
        &      &    &     \\
\hline  
BP1 & 155.56 & 136.11 & 51.85 & 14.19 & 11.64  & 3.01 \\ 
BP2 & 4202.42 & 3262.38 & 1334.47 & 321.74 & 256.86  & 58.14 \\
BP3 & 835.49 & 686.61 & 277.70 & 70.70 & 57.26 & 13.76  \\
BP4 & 126.93 & 126.70 & 79.36 & 12.06 & 8.22 & 0.88\\
BP5 & 93.77 & 88.93 & 55.13& 9.16&6.23 & 0.78 \\
BP6 & 29.66 & 23.81 & 11.58 & 2.58 & 2.13 & 0.58 \\ 
BP7 & 364.38 & 308.97 & 126.35 & 27.61 & 20.34 & 5.15\\
BP8 & 95.58  & 71.47 & 30.46& 7.48 & 6.20 & 1.63 \\
BP9 & 731.08 & 650.39 & 241.20 & 69.16 & 54.65 & 13.63   \\
BP10 & 29.60 & 27.11 & 10.32 & 2.91 & 2.32 & 0.60 \\
\hline 
\end{tabular}
\caption{The cut-flow table for the (mono-jet $+\met$) final state, showing the change in signal 
cross-sections for the ten different benchmark points. The cuts ({\bf C0, D1 -- D5}) are defined in the text in 
Sec. \ref {sec:monojet}.} 
\label{signals2NLO} 
\end{center}
\end{table} 
\begin{itemize}
\item {\bf D4}: We demand $\met > 160$ GeV. In our case, this decreases the SM background substantially, as opposed to the SUSY 
signals (see Table~\ref{signals2NLO} and ~\ref{backgrounds2}).  
\end{itemize}
 We also find that a hard effective mass ($M_{Eff}$) cut is also quite efficient in 
 suppressing the SM background as compared to SUSY signal events for 
 the mono-jet+$\met$ channel. 
\begin{itemize}
\item {\bf D5}: We set $M_{Eff} > 800$ GeV for the analysis which again helps to remove the 
huge QCD background as well as reduce the other dominant contributions. Although the signal events 
are also reduced considerably, the signal-to-background ratio improves significantly after the 
$M_{Eff}$ cut.
\end{itemize}
\begin{table}[h]
\small
\begin{center}
\footnotesize
\begin{tabular}{|c|c|c|c|c|c|c|c|}    
\hline
\multicolumn{2}{|c|}{SM Backgrounds} &
\multicolumn{5}{|c|}{Effective cross-section after the cuts (in pb)} \\
\hline
Channels & Production                    
&{\bf C0 $ + $ D1} & {\bf D2} & {\bf D3} & {\bf D4} &  {\bf D5}    \\
       &  (in pb)  & &      
        &      &    &           \\
\hline 
$\rm t \bar{t}$ +${\leq}$ 2 jets & 722.94 & 573.89 & 171.12 & 21.52 & 2.135 & 0.119\\
t +${\leq}$ 3 jets & 330.57 & 278.05 & 41.14 & 6.17 & 0.355   & 0.011   \\
QCD(${\leq}$4 jets)& 2E+08 & 7.6E+07 & 417461 & 46034 & 2584 & ${\sim}$0\\  
Z +${\leq}$ 4 jets & 57088 & 18924.1 & 446.41 & 52.25 & 6.66  & 0.255  \\
W + ${\leq}$4 jets &  197271 & 50478.5  & 1167.56 & 139.332 & 8.98   & 0.534    \\
WZ + ${\leq}$2 jets & 53.8 & 37.92 & 6.896 & 0.953 & 0.208 & 0.0161 \\
ZZ + ${\leq}$2 jets & 13.69 & 9.77 & 1.03 & 0.158 & 0.0498  & 0.00264 \\

\hline 
Total & \multicolumn{5}{|c|}{} &  \\
background & \multicolumn{5}{|c|}{} & 0.938 \\
\hline 
\end{tabular}
\caption{
The cut-flow table for the (mono-jet $+\met$) final state, showing the change in cross-sections for 
the different subprocesses contributing to the SM background. The cuts ({\bf C0, D1 -- D5}) are defined in 
the text in Sec. \ref {sec:monojet}.
}
\label{backgrounds2} 
\end{center}
\end{table} 

Tables~\ref{signals2NLO} and ~\ref{backgrounds2} summarise the effect of the cuts ({\bf C0, D1 -- D5}) 
on the SUSY signals and SM background cross-sections respectively. 
For both signal and background, we have used the NLO cross-sections as before.
It is clear from Tables~\ref{signals2NLO} and \ref{backgrounds2} that our choice of cuts 
for the mono-jet $+\met$ final state, although quite helpful in suppressing the SM background to 
improve the signal significance is however not an improvement over the multi-jet 
$+\met$ channel. We show the significance of the signal for all the benchmark points in the 
mono-jet $+\met$ channel in Table~\ref{mono-significanceNLO} with the same integrated luminosity 
(${\mathcal L}=100, 500$ and $1000$ fb$^{-1}$).
\begin{table}[t]
\small
\begin{center}
\begin{tabular}{|c|c||c|c|c||c|c|c|}
\cline{3-7}
\multicolumn{1}{c}{} &
\multicolumn{1}{c||}{}&
\multicolumn{3}{|c||}{{Statistical significance}} & \multicolumn{2}{|c|}{Required Luminosity} \\
\multicolumn{1}{c}{} &
\multicolumn{1}{c||}{}&
\multicolumn{3}{|c||}{{(${\mathcal S}$)}} & \multicolumn{2}{|c|}{(in $fb^{-1}$)} \\
\hline
Signal & $ \rm m_{\widetilde{\chi}_1^0}$({\rm GeV}) & 
${\mathcal L}=100 $  & 
${\mathcal L}=500 $  & ${\mathcal L}=1000 $  & ${\mathcal S}=3{\sigma}$ & ${\mathcal S}=5{\sigma}$  \\ 
\cline{1-7}
BP1 & 1406.4 
& 0.98 
& 2.19 & 3.10 & 937.11 & 2603.08 \\ 
BP2 & 842.4 
&18.98 &
42.44 & 60.02 & 2.50 & 6.94  \\
BP3 & 1096.3  
&4.49 & 
10.03 & 14.20 & 44.64 & 124.00 \\
BP4 & 1323.9  
&0.29 
& 0.64 & 0.91 & 10926.44 & 30351.22 \\ 
BP5 & 1417.6 
&0.25  
&0.57 & 0.81 &14400 & 40000 \\ 
BP6 & 1862.0 
&0.19  
&0.42 & 0.60 &  24930.75 & 69252.08  \\ 
BP7 & 1189.0 
&1.68 &
3.76 & 5.31 & 318.88 & 885.77 \\
BP8 & 1496.3
 &0.53 & 
1.19 & 1.68 &  3203.99 & 8899.96\\
BP9 & 1095.4 
&4.45&
9.95 & 14.07 & 45.44 & 126.25\\ 
BP10 & 1709.3 &
0.20 &
0.44 & 0.62 & 22500 & 62500 \\
 \hline
\end{tabular}
\caption{
Statistical significance of the signal for different benchmark points in the 
mono-jet $+\met$ analysis at 13 TeV LHC.  The significance is estimated for three values of integrated luminosity (${\mathcal L}=100, 500$ and $1000$ fb$^{-1}$). We also estimate the required integrated 
luminosity to achieve a $3\sigma$ and $5\sigma$ excess for each benchmark point at LHC 
with $\sqrt{s}=13$ TeV.
}
\label{mono-significanceNLO}
\end{center}
\end{table}

For the mono-jet +$\met$ channel too, we find that the lightest spectrum, {\tt BP2} will be discovered at the earliest.
The heavier spectra, {\tt BP6 and BP10} are no more better observable in the mono-jet +$\met$ channel as they were in 
the multi-jet +$\met$ channel. 
Among others, large number of signal spectra such as, {\tt BP4, BP5, BP8}, have low significances even at 
1000 fb$^{-1}$ whereas {\tt BP3, BP9, BP7 and BP1} may be observed at moderate ($\sim$45 fb$^{-1}$) to high 
($\sim$1000 fb$^{-1}$) luminosities at the LHC. It is important to note that the squark-gluino masses and 
hierarchy dictate the hardness of the cascade jets. As per our selection criteria, {\bf D2} rejects events 
with additional hard jets while retaining many more with softer accompanying jets, thereby enhancing the 
significance in general. However, it adversely affects cases such as {\tt BP4} which have larger mass gaps. 

Thus, although the multi-jet + $\met$ channel provides increased signal significances for all the 
benchmark points, mono-jet + $\met$ channel still remains a viable window for observing compressed spectra. 
Overall, the efficacy of both these channels depends on the splitting among the LSP,
lighter stop, gluino and first two generation squark masses. 
The benchmark points where the masses of the first two generation squarks and the gluino are separated from the LSP 
by about 50 GeV at most are found to have better signal to background ratio in the multi-jet +$\met$ final state 
when compared to mono-jet +$\met$. However, two spectra with similar $\wt q$-$\wt g$ masses
resulting in similar production cross-sections, are expected to differ in their relative 
sensitivities to the two final states depending upon the $\wt q$-$\wt g$-$\lspone$ mass gaps. 
Let us consider {\tt BP5} and {\tt BP8} for example. Although the $\wt q$'s and the $\wt g$ masses 
are very similar, $\Delta M_i$ in {\tt BP8} is much smaller than that in {\tt BP5} 
because of their different LSP masses. Naturally, {\tt BP5} provides a better signal 
significance than {\tt BP8} when multi-jet + $\met$ final state is considered but the situation is reversed when 
we do a mono-jet + $\met$  analysis. {\tt BP3} and {\tt BP9} despite having similar LSP mass, are different in 
terms of the $\wt q$-$\wt g$ mass hierarchy. {\tt BP3}, as a consequence of having smaller gluino 
mass, has a larger production cross-section, but due to the presence of more number of softer jets in 
{\tt BP9}, it does slightly better than {\tt BP3} in terms of signal significance in the mono-jet analysis. 
{\tt BP4} despite having smaller production cross-section than {\tt BP1}, has a better signal significance for 
multi-jet + $\met$  final state due to the presence of more number of harder jets. On the other hand, {\tt BP1} 
does better if mono-jet + $\met$ final state is considered. {\tt BP2} prevails over all the other benchmark points in 
terms of signal significance in both the final states due to its large production cross-section.
{\tt BP6} and {\tt BP10} having very heavy $\wt q$-$\wt g$ spectrum, are unlikely to be probed even at high luminosities.    
\section{Summary and conclusion}
\label{sec:summary}
In this work, we have considered the compressed SUSY scenario within the phenomenological MSSM framework 
that is consistent with all the present collider and DM data. We observe that achieving a 
substantial compression in the whole SUSY spectrum while being consistent with the observed Higgs boson mass 
requires relatively heavy masses for the sparticles. Since at least one of the stop masses needs to be 
heavy (above TeV) in order to enhance the lightest CP-even Higgs boson mass to the allowed range, better 
compression in the parameter space is obtained as we consider heavier LSP masses which nonetheless address the naturalness problem. 
Such mass ranges, we emphasize, are beyond the reach of the 8 TeV run, and therefore, warrant a close 
investigation in the context of 13/14 TeV LHC.
We observe that having a large $\mu$-parameter, too, can achieve tighter compression in the 
remaining spectrum. 

We select ten representative benchmark points 
from the currently allowed parameter space with all kinds of mass hierarchies and explore their detection possibility 
at the 13 TeV run of the LHC. Similar results can be expected if the upgradation to 14 TeV takes place. 
We analyse both the conventional multi-jet +$\met$ channel and the mono-jet +$\met$ channel. 
We observe that although mono-jet +{\met} channel may be a viable option for this kind of scenario, 
a multi-jet +$\met$ final state provides better statistical significance over the SM background for all 
our benchmark points.
\section*{Acknowledgements}
The work of JD, SM, BM and SKR is partially supported by funding available from the Department of Atomic Energy, Government of India, for the Regional 
Centre for Accelerator-based Particle Physics (RECAPP), Harish-Chandra Research Institute. 
PK thanks RECAPP for hospitality during this work.
The authors would like to thank Olivier 
Mattelaer for some useful suggestions regarding Madgraph. 
We also thank S. Banerjee, J. Beuria, A. Choudhury, U. Maitra and T. Mandal for 
useful discussions. Computational work for this study was carried out at the 
cluster computing facility in the Harish-Chandra Research Institute (http://www.hri.res.in/cluster). 
\appendix 
\label{sec:appendix}

\providecommand{\href}[2]{#2}
\addcontentsline{toc}{section}{References}
\bibliographystyle{JHEP}
\bibliography{compsusy}

\end{document}